\documentclass[twocolumn,aps,prd,preprintnumbers,showpacs,superscriptaddress,nofootinbib,amsmath,amssymb,floats,floatfix,showkeys,notitlepage,longbibliography]{revtex4-1}
\usepackage{comment}
\usepackage{graphicx}
\usepackage{xcolor}
\usepackage{longtable}
\usepackage{colortbl} 
\usepackage{subfigure}
\usepackage{palatino}
\usepackage[commandnameprefix=always]{changes}
\usepackage{hyperref}
\hypersetup{colorlinks=true,linkcolor=blue,urlcolor=blue,citecolor=blue}
\usepackage[toc,page]{appendix}
\usepackage[normalem]{ulem}

\usepackage{epsfig}
\usepackage{epstopdf}
\usepackage{epstopdf}

\usepackage{orcidlink}
\usepackage{lipsum}
\usepackage{graphicx}
\usepackage{subfigure}
\usepackage{palatino}
\usepackage{sans}
\usepackage{adjustbox}
\usepackage{latexsym}
\usepackage{amsmath}
\usepackage{amssymb}
\usepackage{amsfonts}
\usepackage{dcolumn}
\usepackage{bm}
\usepackage{tikz}
\usepackage{bigints}
\usepackage{array,tabularx,multirow,booktabs}
\usepackage[tracking=true]{microtype}
\SetTracking{}{500}
\SetTracking{encoding={*}, shape=sc}{40}
\UseRawInputEncoding %for inputenc error%
\allowdisplaybreaks
\usepackage{adjustbox}
\usepackage{latexsym}
\usepackage{amsmath}
\usepackage{amssymb}
\usepackage{amsfonts}
\usepackage{dcolumn}
\usepackage{bm}
\usepackage{tikz}
\usepackage{bigints}
\usepackage{array,tabularx,multirow,booktabs}
\usepackage[tracking=true]{microtype}
\usepackage{color}
\UseRawInputEncoding %for inputenc error%
\allowdisplaybreaks

\begin{document} \sloppy

%\title{Dynamics, Quasi-Periodic Oscillations, and Thermodynamic Phase Structure of Schwarzschild-Asymptotic Magnetic Black Holes in Confining Nonlinear Electrodynamics}

\title{Confining nonlinear electrodynamics black holes: from thermodynamic phases to high-frequency phenomena with accretion process}

%\title{Confining nonlinear electrodynamics black holes: from thermodynamic phases to high-frequency QPOs}

\author{Erdem Sucu
\orcidlink{0009-0000-3619-1492}
}
\email{erdemsc07@gmail.com}
\affiliation{Department of Physics, Eastern Mediterranean
University, Famagusta, 99628 North Cyprus, via Mersin 10, Turkiye}

\author{\.{I}zzet Sakall{\i}
\orcidlink{0000-0001-7827-9476}}
\email{izzet.sakalli@emu.edu.tr}
\affiliation{Department of Physics, Eastern Mediterranean
University, Famagusta, 99628 North Cyprus, via Mersin 10, Turkiye}

\author{Orhan D\"{o}nmez
\orcidlink{0000-0001-9017-2452}}
\email{orhan.donmez@aum.edu.kw}
\affiliation{College of Engineering and Technology, American University of the Middle East, Egaila 54200, Kuwait}

\author{G. Mustafa}
\email{gmustafa3828@gmail.com}
\affiliation{Department of Physics, Zhejiang Normal University, Jinhua, 321004, People’s Republic of China}

\begin{abstract}
We investigate a static, spherically symmetric black hole solution arising from Einstein gravity coupled to a confining nonlinear electrodynamics model that reproduces Maxwell theory in the strong-field regime while introducing confinement-like corrections at large distances. The resulting metric function is asymptotically Schwarzschild but carries a characteristic $Q^3/(9\xi^2 r^4)$ correction, where $Q$ is the magnetic charge and $\xi$ is the nonlinear electrodynamics parameter, with the conventional Reissner-Nordström term $Q^2/r^2$ absent. We analyze the horizon structure and construct three-dimensional embedding diagrams to visualize spatial geometry. Using the Gauss-Bonnet theorem, we compute the weak-field deflection angle in vacuum, cold plasma, and axion-plasmon media, finding that the nonlinear electromagnetic corrections reduce the total bending compared to Schwarzschild at fixed Arnowitt-Deser-Misner mass. The gravitational redshift, Joule-Thomson expansion coefficient, and heat capacity are derived, revealing phase transitions and inversion curves that depend on the model parameters. We obtain closed-form expressions for the photon sphere radius, Lyapunov exponent, and shadow size, demonstrating their sensitivity to $Q$ and $\xi$ along observable Intensities. Fully relativistic hydrodynamical simulations of Bondi-Hoyle-Lyttleton accretion show that the confining geometry produces a $\sim 40\%$ enhancement in mass accretion rate relative to Schwarzschild and generates quasi-periodic oscillations with stable $3\!:\!2$ and $2\!:\!1$ frequency ratios matching observations from black hole X-ray binaries. These results establish the confining nonlinear electrodynamics black hole as a testable model that can reproduce high-frequency quasi-periodic oscillation pairs without invoking black hole spin.
\end{abstract}

\date{\today}


\maketitle
%\tableofcontents
{\color{black}
\section{Introduction} \label{isec1}

Black holes (BHs) represent one of the most striking predictions of general relativity (GR), serving as natural laboratories for testing gravitational physics in the strong-field regime \cite{will2014confrontation,Berti:2015itd,Johannsen:2016uoh}. Since the direct detection of gravitational waves from BH mergers by LIGO/Virgo \cite{abbott2016observation,abbott2016tests} and the imaging of BH shadows by the Event Horizon Telescope (EHT) collaboration \cite{EventHorizonTelescope:2019dse,akiyama2022first}, observational constraints on the properties of BH have reached unprecedented precision. These advances have opened new opportunities to probe deviations from the Schwarzschild and Kerr solutions predicted by GR, including modifications arising from alternative theories of gravity, quantum corrections, and coupling to nonlinear matter fields \cite{bonanno2000renormalization,Born:1934gh,Ayon-Beato:1999qin}.

Nonlinear electrodynamics (NED) provides a natural extension of Maxwell theory that becomes relevant in the presence of strong electromagnetic fields, such as those expected near magnetized compact objects \cite{heisenberg2006consequences,lai2001matter}. Classical NED models, including Born-Infeld \cite{Born:1934gh} and Euler-Heisenberg \cite{heisenberg2006consequences} theories, were originally introduced to regularize the self-energy of point charges and to incorporate quantum electrodynamic corrections, respectively. When coupled with GR, these theories yield modified BH solutions whose properties differ from standard Reissner-Nordström (RN) geometry in ways that can be tested through gravitational lensing, shadow observations, and quasi-periodic oscillation (QPO) measurements \cite{ayon1998regular,hendi2014thermodynamic}. A distinctive class of NED models has recently attracted attention: those exhibiting confinement-like behavior, where Maxwell electrodynamics is recovered in the strong-field limit while nonlinear corrections dominate at large distances \cite{sucu2025dynamics}. This inverted structure—opposite to conventional NED theories—produces BH spacetimes that are asymptotically Schwarzschild, yet carry nontrivial electromagnetic signatures in the near-horizon region.

The thermodynamic properties of BHs have been studied extensively since the seminal work of Bekenstein and Hawking established the connection between horizon area, entropy, and thermal radiation \cite{hawking1974black,bekenstein1973black,gursel2025thermodynamics,WOS:001565141800002NPB}. In the extended phase space formalism, where the cosmological constant is treated as a thermodynamic pressure, BHs exhibit phase transitions analogous to those in ordinary thermodynamic systems \cite{henneaux1984cosmological}. The Joule-Thomson expansion (JTE), which describes isenthalpic processes where enthalpy remains constant while pressure varies, has been applied to BH systems to characterize cooling and heating regimes \cite{sucu2025quantumOzcan,aydiner2025regular,Mustafa:2024fau,Javed:2025dit}. The heat capacity determines thermal stability: BHs with positive heat capacity can equilibrate with a thermal bath, while those with negative heat capacity undergo runaway evaporation \cite{sahan2025quantum,york1986black,davies1989thermodynamic}. These thermodynamic quantities depend sensitively on the underlying metric function, making them useful probes of modifications to GR.

Gravitational lensing—the bending of light by massive objects—serves as one of the cleanest tests of BH geometry \cite{Gibbons:2008rj,sucu2025astrophysical}. The weak-field deflection angle can be calculated using the Gauss-Bonnet theorem (GBT), a topological method that relates local curvature to global geometric properties \cite{Gibbons:2008rj}. In astrophysical environments, photons propagate through dispersive media such as ionized plasma or dark matter (DM) halos, which introduce frequency-dependent corrections to the deflection \cite{sucu2025charged}. The potential coupling between axions—hypothetical particles motivated by the strong CP problem and string theory—and photons adds further modifications that could provide observational signatures of axionic DM \cite{sucu2024effect}.

The photon sphere, consisting of unstable circular null geodesics, defines the boundary of the BH shadow and controls the eikonal limit of quasinormal mode (QNM) frequencies \cite{synge1966escape,hioki2009measurement}. The Lyapunov exponent quantifies the instability time scale of these orbits and determines the decay rate of the ringdown signals following the BH mergers \cite{cardoso2009geodesic}. Shadow observations by the EHT have already placed constraints on BH parameters, and next-generation instruments aim to resolve the photon ring substructure with sufficient precision to test strong-field gravity predictions \cite{sucu2025quantumHassan,Mustafa:2024fau,bambi2017testing,falcke1999viewing}.

The QPOs observed in the X-ray emission of accreting BH systems provide another window into strong-field physics \cite{Belloni:2019sot,Zhang:2022epl}. High-frequency QPOs (HFQPOs) in the $50$--$450$ Hz band, often appearing in $3\!:\!2$ or $2\!:\!1$ frequency ratios, have been detected in microquasars such as GRS~1915+105, GRO~J1655-40, and XTE~J1550-564 \cite{Wang:2008xda,Stuchlik:2016brp,Chakrabarti:2005bd}. These commensurabilities are typically interpreted within epicyclic resonance models that invoke the spin of Kerr BHs. However, alternative mechanisms based on modified gravity or nonlinear electromagnetic effects could reproduce similar frequency ratios without requiring frame-dragging, offering testable predictions for non-rotating BH models \cite{bambi2018testing,sucu2025nonlinear,vsramkova2015black}.

The Bondi-Hoyle-Lyttleton (BHL) accretion mechanism describes the capture of matter that flows supersonically towards a compact object, leading to the formation of shock cones in the downstream region \cite{bondi1944mechanism,hoyle1939effect,ruffert1994three}. Numerical simulations of BHL accretion around BHs reveal that the shock cone behaves as a resonant cavity, trapping density and pressure modes that generate QPOs through nonlinear coupling \cite{donmez2011development,donmez2012relativistic,Donmez:2022dze}. The power spectral density (PSD) analysis of the mass accretion rate extracts these oscillation frequencies and enables direct comparison with observational data. The innermost stable circular orbit (ISCO) marks the inner boundary of stable accretion flows and sets the characteristic frequencies of the system \cite{donmez2024bondi,van1989fourier}.

Motivated by these developments, in the present work we investigate a confining NED BH whose metric function approaches the Schwarzschild form at large distances but carries a $Q^3/(9\xi^2 r^4)$ correction arising from the nonlinear electromagnetic sector, where $Q$ denotes the magnetic charge and $\xi$ is the NED parameter. Unlike the standard RN solution, this geometry lacks the $Q^2/r^2$ Coulomb term, reflecting the confining nature of the electromagnetic field. The Arnowitt-Deser-Misner (ADM) mass receives a logarithmic contribution from the electromagnetic self-energy, introducing nontrivial parameter dependence into all observable quantities. Our aims are threefold: (i) characterizing the horizon structure, lensing properties, and thermodynamic phase behavior of this BH model; (ii) computing the photon sphere, shadow, and Lyapunov exponent that govern QNM frequencies; and (iii) performing fully relativistic hydrodynamical simulations of BHL accretion and extracting the QPO spectrum, comparing with both Schwarzschild predictions and observational data from BH X-ray binaries.

The paper is organized as follows. In Sec.~\ref{isec2}, we derive the spacetime geometry of the confining NED BH, analyze the event horizon (EH) structure for various combinations of $Q$ and $\xi$, and present 3D embedding diagrams that visualize the spatial geometry. Section~\ref{isec3} employs the GBT to compute the weak-field deflection angle in vacuum and in a cold plasma medium, while Sec.~\ref{isec4} extends the analysis to include the axion-photon coupling. The gravitational redshift is examined in Sec.~\ref{isec5}, highlighting the $r^{-4}$ correction characteristic of this model. Section~\ref{isec6} investigates the JTE coefficient and the heat capacity, identifying inversion curves and phase transitions in the extended thermodynamic phase space. The Lyapunov exponent for circular null geodesics is derived in Sec.~\ref{isec7}, followed by the photon sphere and shadow analysis in Sec.~\ref{isec8}. In Sec.~\ref{isec8a} observable intensities of confining NED BH with static spherical accretion are discussed. Sections~\ref{isec9} and \ref{isec10} present numerical simulations of BHL accretion and PSD analysis of the resulting QPO spectrum, comparing non-extremal (NE), extremal, and naked singularity (NS) configurations with the Schwarzschild baseline. Finally, Sec.~\ref{isec11} summarizes our findings and outlines directions for future research. Throughout this work, we adopt geometrized units where $G = c = 1$ unless otherwise stated.

\section{Spacetime Geometry of the Confining NED BH} \label{isec2}

Within the framework of GR minimally coupled to NED, the construction of regular or phenomenologically distinct BH solutions typically requires a judicious choice of the electromagnetic Lagrangian density \cite{ayon1998regular,Kruglov:2015fbl}. In this work, we examine a recently proposed NED model that exhibits a rather unusual property: it reproduces standard Maxwell electrodynamics in the strong-field regime while introducing confinement-like corrections at large distances where the field becomes weak \cite{mazharimousavi2024confinement}. This inverted behavior—opposite to what occurs in classical NED theories such as Born-Infeld or Euler-Heisenberg—renders the model particularly attractive for investigating asymptotically flat spacetimes carrying nontrivial electromagnetic structure.

The gravitational action governing the system under consideration takes the form:
\begin{equation}
S = \int d^4x \, \sqrt{-g} \left( \frac{R}{16\pi} + \mathcal{L}(F) \right),
\end{equation}
where $R$ denotes the Ricci scalar, $g$ represents the metric determinant, and $\mathcal{L}(F)$ stands for the NED Lagrangian density expressed as a function of the electromagnetic invariant $F = \frac{1}{4} F_{\mu\nu} F^{\mu\nu}$. For magnetic configurations, the model assumes the specific structure:
\begin{equation}
\mathcal{L}(F) = -\varepsilon(F) F,
\end{equation}
with a field-dependent dielectric function $\varepsilon(F)$ given by \cite{mazharimousavi2024confinement}:
\begin{equation}
\varepsilon(F) = \frac{16\left[3\sqrt{2F} + \xi\left(\xi + \sqrt{\xi^2 + 4\sqrt{2F}}\,\right)\right]\sqrt{2F}}{3\left(\xi + \sqrt{\xi^2 + 4\sqrt{2F}}\,\right)^4},
\end{equation}
where the parameter $\xi > 0$ controls the strength of the nonlinear corrections and encodes the characteristic of the confinement effect of this model.

The BH solution derived from variational principles admits a static, spherically symmetric line element:
\begin{equation}
ds^2 = -f(r) dt^2 + \frac{dr^2}{f(r)} + r^2 (d\theta^2 + \sin^2\theta\, d\phi^2). \label{metric}
\end{equation}
We consider a configuration of purely magnetic field sourced by a magnetic monopole of charge $Q$. The two-form electromagnetic field strength reads:
\begin{equation}
F = Q \sin\theta\, d\theta \wedge d\phi,
\end{equation}
which automatically satisfies the Bianchi identity $dF = 0$ and the generalized Maxwell equations following the NED Lagrangian. Solving the coupled Einstein-NED field equations yields a metric function $f(r)$ whose asymptotic expansion as $r \to \infty$ takes the form \cite{mazharimousavi2024confinement}:
\begin{equation}
f(r) = 1 - \frac{2M_{\text{ADM}}}{r} + \frac{Q^3}{9\xi^2 r^4} + \mathcal{O}\left(\frac{1}{r^6}\right), \quad \text{as } r \rightarrow \infty.\label{metricFun}
\end{equation}
A striking feature of this expression is the complete absence of the standard RN term $Q^2/r^2$ that normally characterizes charged BHs in GR coupled to Maxwell electrodynamics. This absence stems directly from the confining nature of the NED model: Maxwell theory governs physics only in the strong-field domain (small $r$), whereas nonlinear corrections become dominant at large distances, effectively screening the long-range Coulomb potential. As a result, the geometry approaches the Schwarzschild solution at spatial infinity, modified only by a subleading $1/r^4$ correction that decays significantly faster than the conventional Coulomb term.

The ADM mass receives a contribution from the electromagnetic self-energy stored in the nonlinear field and is expressed as \cite{mazharimousavi2024confinement}:
\begin{equation}
M_{\text{ADM}} = M + \frac{2\sqrt{2}}{3} \xi Q^{3/2} \ln\left(2\xi\sqrt{2Q}\right),
\end{equation}
where $M$ is the bare Schwarzschild mass parameter. The logarithmic dependence on the product $\xi Q^{1/2}$ reflects how nonlinear electromagnetic energy accumulates and contributes to the total gravitating mass measured by distant observers.

Figure~\ref{fig:metric_function} displays the radial behavior of the metric function $f(r)$ for several representative combinations of the magnetic charge $Q$ and the NED parameter $\xi$, with the Schwarzschild mass fixed at $M=1$. The zero crossings of $f(r)$ identify the locations of the EH. When $Q$ increases at fixed $\xi$, the curves shift downward and the system progresses toward extremality or even naked singularity (NS) configurations where no horizon exists. In contrast, raising $\xi$ to the fixed value $Q$ enhances the ADM mass contribution, which tends to restore the horizon structure and pull the spacetime back into the BH regime. The plot clearly distinguishes NE configurations possessing two distinct horizons (inner and outer), extremal cases with degenerate horizons, and NS cases where $f(r)$ remains positive for all $r > 0$.

Table~\ref{tab:NED_horizons} presents a detailed analysis of the horizon structure in a range of values $(Q, \xi)$. For small magnetic charge ($Q = 0.1$), horizons exist even when $\xi = 0$, and increasing $\xi$ pushes the inner horizon closer to the origin while the outer horizon remains near $r \approx 2M$, consistent with the Schwarzschild limit. At larger charges ($Q \geq 0.5$), the situation changes qualitatively: configurations with $\xi = 0$ or very small $\xi$ correspond to NS spacetimes, and a finite threshold value of $\xi$ is required to generate horizons. For instance, at $Q = 1.0$, the transition from NS to BH occurs between $\xi = 0.1$ and $\xi = 0.3$. The table also reveals that the radius of the outer horizon increases substantially with $Q$ and $\xi$—reaching $r_+ \approx 9.4M$ for $(Q, \xi) = (2.0, 1.0)$—highlighting the strong influence of the NED parameter on the global causal structure.

\setlength{\tabcolsep}{12pt}
\renewcommand{\arraystretch}{1.6}
\begin{longtable*}{|c|c|c|c|}
\hline
\rowcolor{orange!50}
\textbf{$Q$} & \textbf{$\xi$} & \textbf{Horizon(s)} & \textbf{Configuration} \\
\hline
\endfirsthead
\hline
\rowcolor{orange!50}
\textbf{$Q$} & \textbf{$\xi$} & \textbf{Horizon(s)} & \textbf{Configuration} \\
\hline
\endhead
0.1 & 0.0 & $[1.0570,\ 1.8100]$ & NE (inner+outer) \\
\hline
0.1 & 0.1 & $[0.1834,\ 1.9840]$ & NE (inner+outer) \\
\hline
0.1 & 0.3 & $[0.0868,\ 1.9760]$ & NE (inner+outer) \\
\hline
0.1 & 0.5 & $[0.0615,\ 1.9760]$ & NE (inner+outer) \\
\hline
0.1 & 0.7 & $[0.0490,\ 1.9800]$ & NE (inner+outer) \\
\hline
0.1 & 1.0 & $[0.0384,\ 1.9930]$ & NE (inner+outer) \\
\hline
0.5 & 0.0 & No horizon & NS \\
\hline
0.5 & 0.1 & No horizon & NS \\
\hline
0.5 & 0.3 & $[0.4771,\ 1.8740]$ & NE (inner+outer) \\
\hline
0.5 & 0.5 & $[0.3210,\ 1.9930]$ & NE (inner+outer) \\
\hline
0.5 & 0.7 & $[0.2457,\ 2.1540]$ & NE (inner+outer) \\
\hline
0.5 & 1.0 & $[0.1826,\ 2.4610]$ & NE (inner+outer) \\
\hline
1.0 & 0.0 & No horizon & NS \\
\hline
1.0 & 0.1 & No horizon & NS \\
\hline
1.0 & 0.3 & $[1.2100,\ 1.6130]$ & NE (inner+outer) \\
\hline
1.0 & 0.5 & $[0.6413,\ 2.2900]$ & NE (inner+outer) \\
\hline
1.0 & 0.7 & $[0.4524,\ 2.8920]$ & NE (inner+outer) \\
\hline
1.0 & 1.0 & $[0.3123,\ 3.9590]$ & NE (inner+outer) \\
\hline
1.5 & 0.3 & No horizon & NS \\
\hline
1.5 & 0.5 & $[0.9011,\ 2.8890]$ & NE (inner+outer) \\
\hline
1.5 & 0.7 & $[0.5997,\ 4.1370]$ & NE (inner+outer) \\
\hline
1.5 & 1.0 & $[0.3990,\ 6.3020]$ & NE (inner+outer) \\
\hline
2.0 & 0.0 & No horizon & NS \\
\hline
2.0 & 0.1 & No horizon & NS \\
\hline
2.0 & 0.5 & $[1.0880,\ 3.7830]$ & NE (inner+outer) \\
\hline
2.0 & 0.7 & $[0.7068,\ 5.8350]$ & NE (inner+outer) \\
\hline
2.0 & 1.0 & $[0.4634,\ 9.3920]$ & NE (inner+outer) \\
\hline
\caption{Horizon structure of confining NED BHs. For each combination of magnetic charge $Q$ and NED parameter $\xi$, the corresponding horizon radii are listed with fixed Schwarzschild mass $M=1$. The ADM mass is given by $M_{\text{ADM}} = M + \frac{2\sqrt{2}}{3} \xi Q^{3/2} \ln(2\xi\sqrt{2Q})$, and the metric function is $f(r) = 1 - 2M_{\text{ADM}}/r + Q^3/(9\xi^2 r^4)$. The absence of the standard RN $Q^2/r^2$ term reflects the confining nature of the NED model, where nonlinear corrections dominate at large distances while Maxwell theory is recovered in the strong-field regime. Here, NE and NS stand for non-extremal black hole and naked singularity, respectively.}
\label{tab:NED_horizons}
\end{longtable*}

\begin{figure*}[htbp]
\centering
\includegraphics[width=0.85\textwidth]{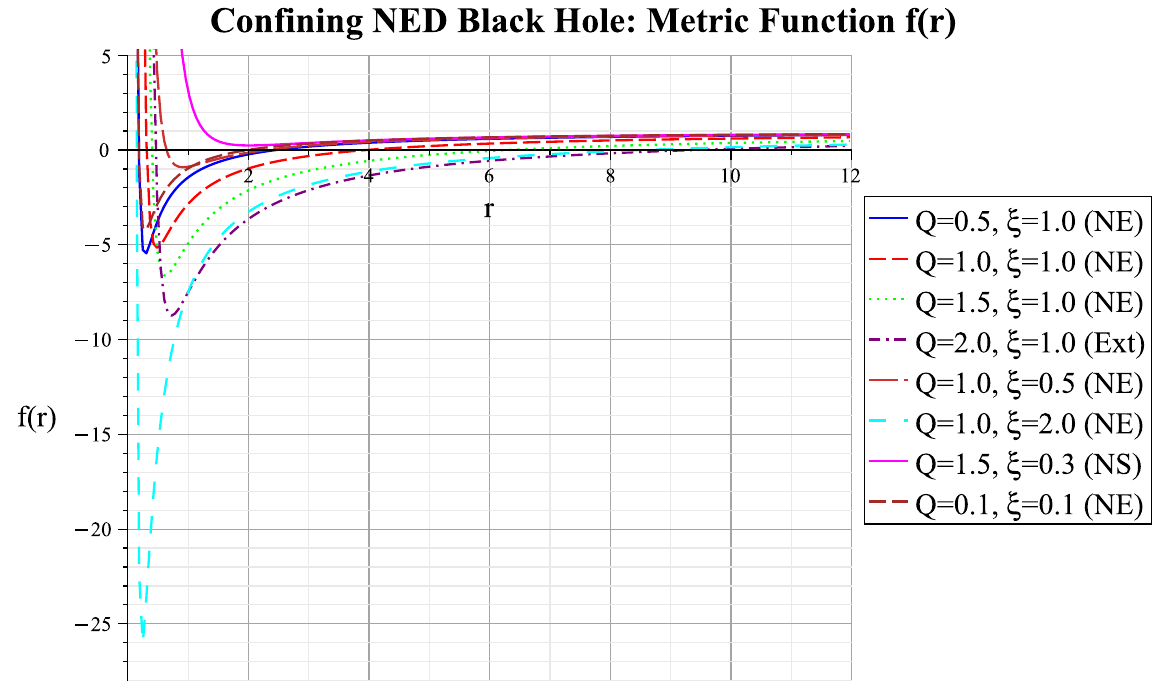}
\caption{Metric function $f(r)$ for the confining NED BH with fixed Schwarzschild mass $M=1$. Representative parameter combinations are displayed with different line styles and colors. The black dotted horizontal line marks $f(r)=0$. Horizon locations correspond to zero crossings, with NE configurations exhibiting two distinct horizons, Ext cases showing degenerate horizons, and NS cases displaying no horizon. The absence of the standard RN $Q^2/r^2$ term results in asymptotically Schwarzschild behavior at large $r$.}
\label{fig:metric_function}
\end{figure*}

The modified spacetime thus admits BH solutions with up to two horizons depending on the interplay among $M$, $Q$, and $\xi$. Although the NED correction to the metric function becomes negligible at large $r$, it plays a decisive role near the EH and at short distances, modifying thermodynamic quantities such as Hawking temperature, heat capacity and phase transition behavior \cite{balart2014regular,fernando2003charged}.

To visualize the spatial geometry of the confining NED BH, we construct isometric embedding diagrams by embedding the spatial slice equatorial ($\theta = \pi/2$) into three-dimensional Euclidean space. The embedding surface is generated by revolving the profile of the metric function around the symmetry axis, with the radial coordinate mapped onto the horizontal plane and the value of $f(r)$ determining the vertical displacement. Figure~\ref{fig:embedding_panels} shows the resulting funnel-shaped geometries for four representative parameter sets.

\begin{figure*}[ht!]
    \centering
        \includegraphics[width=0.35\textwidth]{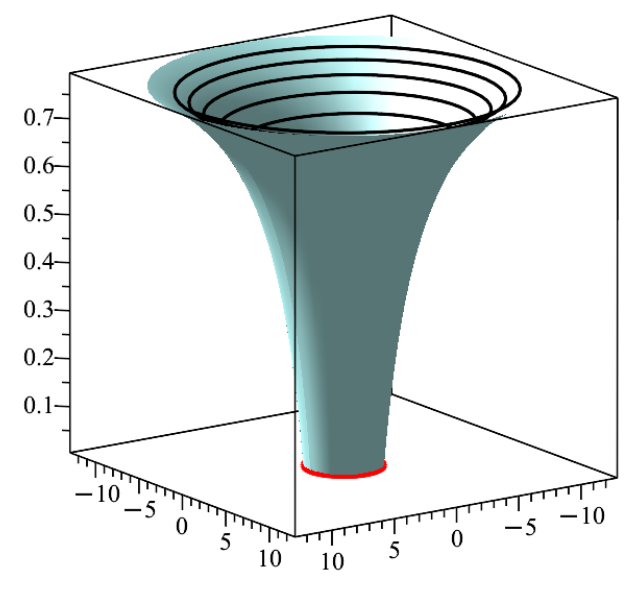}
        \includegraphics[width=0.35\textwidth]{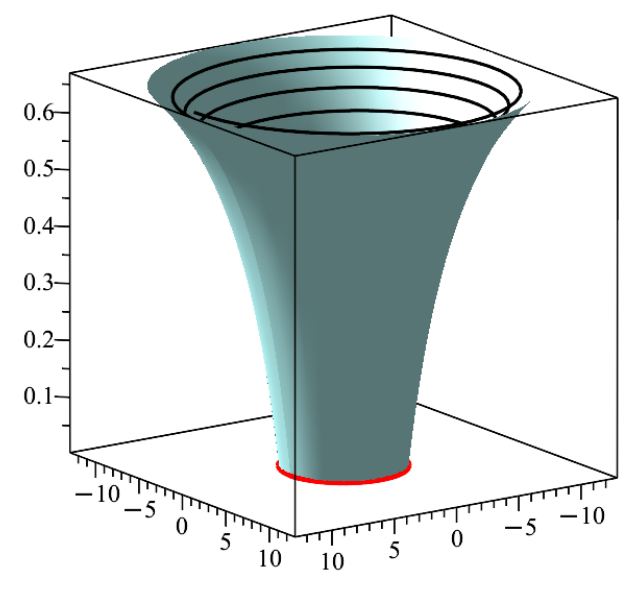}\\
        (i) $Q=0.5$, $\xi=1.0$, $r_H=2.46$ \hspace{1.5cm}
        (ii) $Q=1.0$, $\xi=1.0$, $r_H=3.96$
        \vspace{0.5em}
        
        \includegraphics[width=0.35\textwidth]{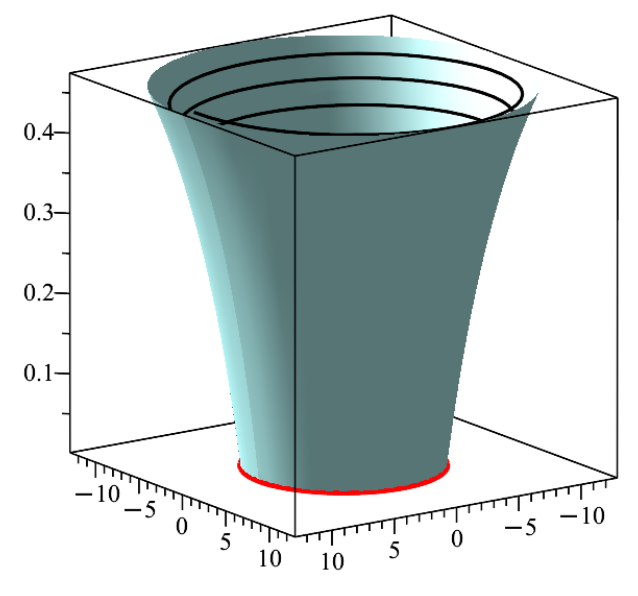}
        \includegraphics[width=0.35\textwidth]{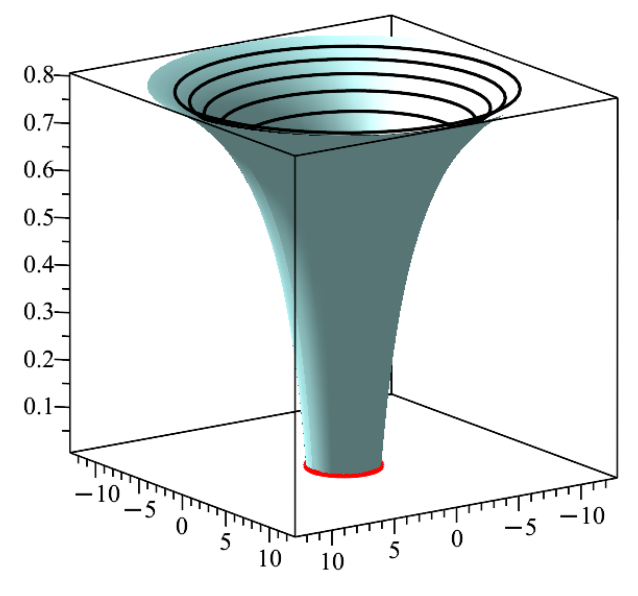}\\
        (iii) $Q=1.5$, $\xi=1.0$, $r_H=6.30$ \hspace{1.5cm}
        (iv) $Q=1.0$, $\xi=0.5$, $r_H=2.29$
\caption{Three-dimensional embedding diagrams of the confining NED BH with metric function $f(r) = 1 - 2M_{\text{ADM}}/r + Q^3/(9\xi^2 r^4)$, where $M = 1$ is fixed and $M_{\text{ADM}} = M + \frac{2\sqrt{2}}{3} \xi Q^{3/2} \ln(2\xi\sqrt{2Q})$. Each panel displays a turquoise surface representing the isometric embedding from the EH $r_H$ to $r = 12$, a black spiral depicting an infalling test particle trajectory, and a red ring marking the EH location. Panels (i)--(iii) illustrate the effect of increasing $Q$ at fixed $\xi=1.0$: larger magnetic charge yields progressively wider throat geometries and larger horizon radii. Comparing panels (ii) and (iv) reveals the influence of $\xi$ at fixed $Q=1.0$: reducing $\xi$ decreases the contribution of the ADM mass and consequently shrinks the radius of the horizon.}
\label{fig:embedding_panels}
\end{figure*}

The embedding diagrams in Fig.~\ref{fig:embedding_panels} expose several key geometric features. Panels (i)--(iii) demonstrate that the EH radius grows monotonically with $Q$ when $\xi$ is held constant at unity: starting from $r_H = 2.46$ at $Q = 0.5$, the horizon expands to $r_H = 3.96$ at $Q = 1.0$ and further to $r_H = 6.30$ at $Q = 1.5$. The gravitational funnel becomes progressively wider and shallower as the horizon size increases. Comparing panels (ii) and (iv) isolates the role of $\xi$ at fixed charge $Q = 1.0$: lowering $\xi$ from $1.0$ to $0.5$ reduces the ADM mass contribution and contracts the horizon from $r_H = 3.96$ to $r_H = 2.29$. The black spiral trajectories illustrate how infalling test particles are captured by the gravitational potential, spiraling inward until they cross the EH marked by the red ring.

In summary, the metric function arising from the NED Lagrangian $\mathcal{L}(F) = -\varepsilon(F)F$ describes a BH spacetime that is asymptotically Schwarzschild yet possesses a nontrivial near-horizon structure shaped by the confining electromagnetic field. This model offers a rare example in which Maxwell electrodynamics governs the strong-field domain while nonlinear effects emerge only in the weak-field limit—precisely the reverse of what occurs in conventional NED theories like Born-Infeld or Euler-Heisenberg. The rich interplay between $Q$ and $\xi$ opens the door to distinct thermodynamic phases and observable signatures that we explore in the following sections.

\section{Weak Gravitational Lensing of the Confining NED BH with Plasma Corrections} \label{isec3}

Gravitational lensing constitutes one of the most powerful observational tools to investigate the geometry of compact objects and to testing BH models in realistic astrophysical environments \cite{vegetti2024strong,guo2022gravitational,ellis2010gravitational,wambsganss1998gravitational,sucu2025probing,sucu2025scalar}. Light rays passing near massive bodies experience deflection due to spacetime curvature, and precise measurements of this bending can discriminate between different gravitational theories and constrain BH parameters \cite{einstein1916foundation,will2014confrontation}. In this section, we investigate the weak deflection of light around the confining NED BH derived in Sec.~\ref{isec2}, first in vacuum and subsequently in a cold, non-magnetized plasma medium that is commonly encountered in astrophysical settings such as accretion flows and galactic halos.

We employ the GBT a topological approach that relates local curvature properties to global geometric invariants—to compute the deflection angle \cite{crisnejo2018weak,crisnejo2019higher}. This method has gained considerable attention in recent years because it offers a coordinate-independent formulation and naturally incorporates finite-distance corrections \cite{WOS:001617915000001}. The starting point is the construction of an optical metric from the null condition $ds^2 = 0$. Restricting motion to the equatorial plane ($\theta = \pi/2$) and using the line element given in Eq.~\eqref{metric}, the optical geometry becomes:
\begin{equation}
dt^2 = \gamma_{ij} dx^i dx^j = \frac{1}{f^2(r)} dr^2 + \frac{r^2}{f(r)} d\phi^2,
\end{equation}
where $\gamma_{ij}$ denotes the two-dimensional spatial metric governing photon trajectories. Introducing a generalized tortoise coordinate through the transformation $dr^* = dr/f(r)$ recasts the optical line element into a conformally flat form:
\begin{equation}
dt^2 = dr^{*2} + \tilde{f}^2(r^*) d\phi^2,\quad \text{with} \quad \tilde{f}(r^*) = \sqrt{r^2/f(r)}.
\end{equation}

The Gaussian curvature $\mathcal{K}$ associated with this optical geometry encodes the gravitational influence of the BH on photon propagation. For the confining NED metric function given in Eq.~\eqref{metricFun}, the curvature is evaluated to \cite{sucu2025charged}:
\begin{multline}
\mathcal{K} = \frac{R}{2}=\frac{2 M_{\text{ADM}}}{r^{3}}+\frac{10 Q^{3}}{9 \xi  r^{6}}+\frac{3 M_{\text{ADM}}^{2}}{r^{4}}\\-\frac{2 M_{\text{ADM}} \,Q^{3}}{r^{7} \xi}+\frac{2 Q^{6}}{27 \xi^{2} r^{10}},
\end{multline}
where the terms proportional to $Q^3$ and $Q^6$ arise specifically from the NED corrections and vanish in the Schwarzschild limit ($Q \to 0$). The leading $1/r^3$ term reproduces the standard GR result, while the higher-order contributions encode the characteristic fingerprint of the confining electromagnetic structure.

Applying the GBT over a domain $\tilde{D}$ bounded by the photon trajectory and a circular arc at large radius, and recognizing that the light ray follows a geodesic of the optical metric, the deflection angle $\Theta$ satisfies the integral relation:
\begin{equation}
\iint_{\tilde{D}} \mathcal{K} \, dS + \int_{0}^{\pi + \Theta} d\phi = \pi.
\end{equation}
Rearranging and evaluating the surface integral yields the total deflection:
\begin{equation}
\Theta = - \int_0^\pi \int_{b/\sin\phi}^\infty \mathcal{K} \, dS,
\end{equation}
where $b$ denotes the impact parameter of the incoming photon. The optical surface element takes the approximate form:
\begin{equation}
dS = \frac{r}{f^{3/2}(r)} dr \, d\phi.
\end{equation}

Carrying out the integration to the relevant order in $M_{\text{ADM}}/b$ and $Q/b$, the vacuum deflection angle becomes:
\begin{multline}
    \Theta =\frac{4 M_{\text{ADM}}}{b}+\frac{3 M_{\text{ADM}}^{2} \pi}{4 b^{2}}-\frac{4 M_{\text{ADM}}^{3}}{b^{3}}\\-\frac{5 Q^{3} \pi}{48 \xi  b^{4}}-\frac{64 M_{\text{ADM}} \,Q^{3}}{225 \xi  b^{5}}+\frac{5 M_{\text{ADM}}^{2} Q^{3} \pi}{16 \xi  b^{6}}\\-\frac{35 Q^{6} \pi}{13824 \xi^{2} b^{8}}-\frac{512 Q^{6} M_{\text{ADM}}}{25515 \xi^{2} b^{9}}.
\end{multline}
The first three terms correspond to the well-known post-Newtonian expansion for Schwarzschild geometry: the leading $4M_{\text{ADM}}/b$ recovers Einstein's classical result, while the $\pi M_{\text{ADM}}^2/b^2$ and $M_{\text{ADM}}^3/b^3$ corrections account for higher-order strong-field effects. The remaining terms—scaling as $Q^3/(\xi b^4)$, $Q^3/(\xi b^5)$, etc., so forth—constitute the modifications induced by NED. Notably, all NED corrections carry negative signs, indicating that the confining electromagnetic field tends to reduce the total bending angle compared to the pure Schwarzschild case at a fixed ADM mass.

Figure~\ref{lens} displays the deflection angle $\Theta$ as a function of the impact parameter $b$ (expressed through the horizon radius $r_h$) and the NED parameter $\xi$, with the Schwarzschild mass and magnetic charge fixed at $M=1$ and $Q=0.5$, respectively. The surface plot reveals that $\Theta$ decreases monotonically as $b$ increases, consistent with the expected $1/b$ over large distances. For fixed impact parameter, increasing $\xi$ produces a mild enhancement of the deflection angle because a larger $\xi$ increases the ADM mass through the logarithmic correction term. The smooth variation across the parameter space confirms the regularity of the optical geometry outside the EH.

\begin{figure}[t]
    \centering
    \includegraphics[width=0.5\textwidth]{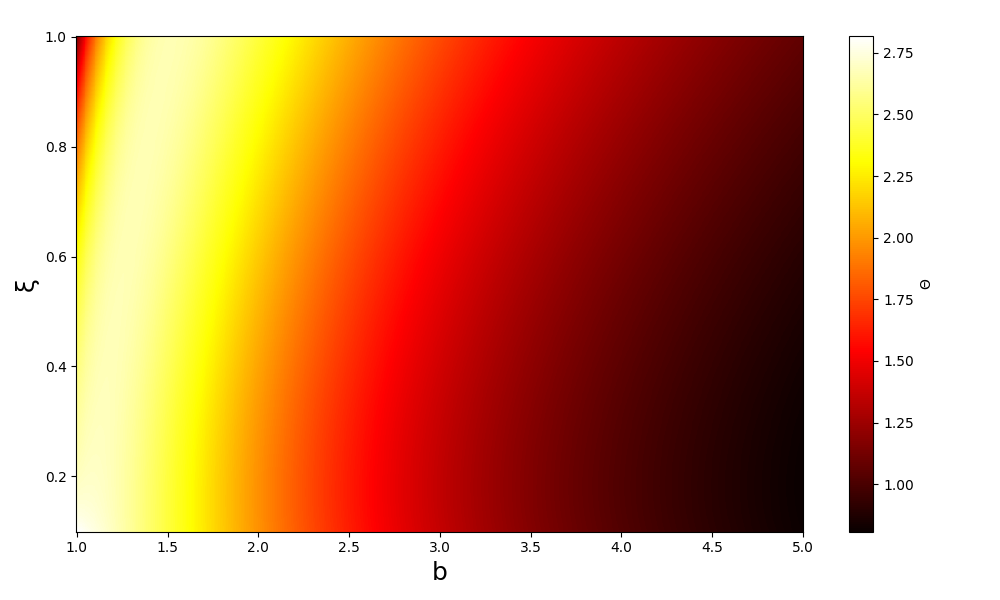}
    \caption{Deflection angle $\Theta$ as a function of the impact parameter $b$ and the NED parameter $\xi$, for fixed values $M=1$ and $Q=0.5$. Larger $\xi$ increases the ADM mass and slightly enhances the bending, while increasing $b$ reduces $\Theta$ following the characteristic inverse power-law behavior.}
    \label{lens}
\end{figure}

In realistic astrophysical environments, photons propagate not through a perfect vacuum but through dispersive media such as ionized gas, accretion disks, or intergalactic plasma \cite{sucu2025quantumRoyal,perlick2002ray}. The presence of plasma modifies the effective refractive index experienced by electromagnetic waves and consequently alters the lensing properties. To incorporate these effects, we introduce a frequency-dependent refractive index \cite{sucu2025astrophysical}:
\begin{equation}
n(r) = \sqrt{1 - \frac{\omega_p^2(r)}{\omega_0^2} f(r)},
\end{equation}
where $\omega_p(r)$ denotes the local plasma frequency determined by the electron density and $\omega_0$ represents the photon frequency measured at spatial infinity. For a homogeneous plasma distribution, the ratio $\omega_p^2/\omega_0^2$ remains constant along the photon path, simplifying the analysis while retaining the essential dispersive physics.

The optical line element in the presence of a plasma becomes:
\begin{equation}
dt^2 = n^2(r) \left[ \frac{1}{f^2(r)} dr^2 + \frac{r^2}{f(r)} d\phi^2 \right],
\end{equation}
and the associated Gaussian curvature transforms to:
\begin{equation}
\tilde{\mathcal{K}} = \frac{\mathcal{R}_{r\phi r\phi}}{n^4(r) r^2 / f^3(r)}.
\end{equation}
Expanding to leading order in both the gravitational potential and the plasma parameter, we obtain:
\begin{multline}
    \tilde{\mathcal{K}}\approx \frac{12 M_{\text{ADM}}^{3} \delta}{r^{5}}+\frac{12 M_{\text{ADM}}^{2} \delta}{r^{4}}+\frac{3 M_{\text{ADM}}^{2}}{r^{4}}-\frac{3 M_{\text{ADM}} \delta}{r^{3}}\\-\frac{2 M_{\text{ADM}}}{r^{3}}+\frac{2 Q^{3} \delta}{\xi  r^{6}}+\frac{10 Q^{3}}{9 \xi  r^{6}}+\mathcal{O}(r^{-7}),
\end{multline}
where we have introduced the dimensionless plasma parameter:
\begin{equation}
    \delta = \frac{\omega_e^2}{\omega_0^2},
\end{equation}
with $\omega_e$ being the electron plasma frequency. The modified area element reads:
\begin{equation}
dS = \left( r - \frac{\omega_p^2}{\omega_0^2} \right) dr \, d\phi,
\end{equation}
and the plasma-corrected deflection angle follows from:
\begin{equation}
\tilde{\Theta} = - \int_0^{\pi} \int_{b/\sin\phi}^\infty \tilde{\mathcal{K}} \, dS.
\end{equation}

Evaluating this integral yields the full expression:
\begin{multline}
    \tilde{\Theta}\approx \frac{4 M_{\text{ADM}}}{b}+\frac{6 M_{\text{ADM}} \delta}{b}+\frac{3 M_{\text{ADM}}^{2} \pi}{4 b^{2}}-\frac{3 M_{\text{ADM}}^{2} \pi  \delta}{4  b^{2}}\\-\frac{4 M_{\text{ADM}}^{3}}{b^{3}}-\frac{32 M_{\text{ADM}}^{3} \delta}{3  b^{3}}+\frac{27 M_{\text{ADM}}^{4} \pi  \delta }{8  b^{4}}\\-\frac{5 \pi  Q^{3}}{48 \xi  b^{4}}-\frac{3 \pi  Q^{3} \delta}{16  \xi  b^{4}}-\frac{32 Q^{3} M_{\text{ADM}}}{45 \xi  b^{5}}-\frac{32 Q^{3} \delta M_{\text{ADM}}}{25  \xi  b^{5}}.
\end{multline}
Several features of this result deserve attention. First, the leading-order plasma correction $6 M_{\text{ADM}} \delta / b$ adds positively to the vacuum term $4 M_{\text{ADM}} / b$, implying that a dispersive medium enhances the deflection for low-frequency photons ($\delta \to 1$) while high-frequency radiation ($\delta \to 0$) recovers the vacuum limit. Second, the cross-term coupling $Q^3$ with $\delta$ demonstrates that plasma effects and NED corrections do not simply superpose, but interact in a nontrivial manner. Third, the alternating signs of successive terms indicate a delicate interplay between gravitational focusing and dispersive refraction.

Figure~\ref{lensplasma} presents the plasma-corrected deflection angle $\tilde{\Theta}$ as a function of the impact parameter and the plasma parameter $\delta$, with $M=1$, $\xi=1$, and $Q=0.5$. The surface exhibits a pronounced increase in $\tilde{\Theta}$ as $\delta$ grows, confirming that denser plasma environments (or lower photon frequencies) produce stronger bending. At fixed $\delta$, the deflection still decreases with increasing $b$, but the rate of decrease is modulated by the plasma content. These results suggest that multi-frequency observations of lensing events near magnetized compact objects could, in principle, disentangle the gravitational contribution from plasma-induced chromatic effects and thereby constrain the NED parameters \cite{bisnovatyi2010gravitational}.

\begin{figure}[t]
    \centering
    \includegraphics[width=0.5\textwidth]{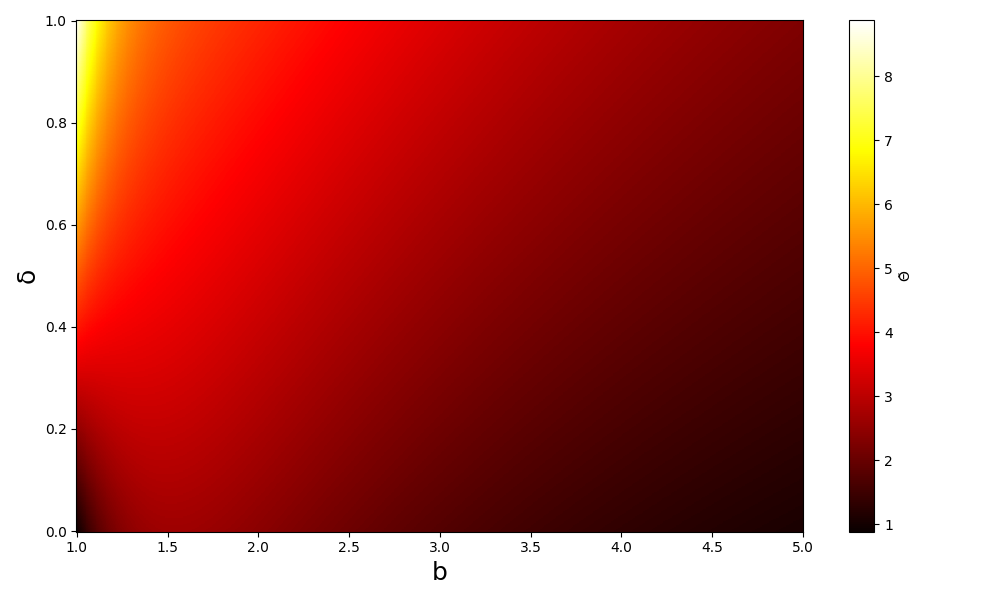}
    \caption{Plasma-corrected deflection angle $\tilde{\Theta}$ as a function of the impact parameter $b$ and the plasma parameter $\delta$, for fixed values $M=1$, $\xi=1$, and $Q=0.5$. Increasing $\delta$ (corresponding to denser plasma or lower photon frequency) enhances the total bending, while the characteristic $1/b$ decay persists at large impact parameters.}
    \label{lensplasma}
\end{figure}

In summary, the weak-field gravitational lensing analysis reveals that the confining NED BH produces a deflection angle that differs from both the Schwarzschild and standard RN results. The NED corrections introduce additional inverse power-law terms scaling with $Q^3/\xi$ and $Q^6/\xi^2$, which become increasingly important at small impact parameters. When plasma effects are included, the deflection acquires a characteristic frequency dependence that could serve as an observational discriminant. These lensing signatures complement the horizon structure and embedding geometry discussed in the previous section, offering an independent avenue for testing the confining NED model against astrophysical data.

\section{Light Deflection in an Axion-Plasmon Medium} \label{isec4}

Axions represent one of the most compelling DM candidates, originally proposed to resolve the strong CP problem in quantum chromodynamics and subsequently identified as a natural consequence of string theory compactifications \cite{Mendonca:2019eke,Wilczek:1987mv}. The coupling between axions and photons, though extremely weak under laboratory conditions, can become observationally significant in the vicinity of compact objects where strong gravitational and magnetic fields coexist \cite{raffelt1988mixing,lecce2025probing}. In this section, we extend the gravitational lensing analysis of the confining NED BH by incorporating axion-photon interactions within a magnetized plasma environment, building upon the vacuum and pure plasma results derived in Sec.~\ref{isec3}.

The theoretical framework for axion electrodynamics augments the standard Maxwell Lagrangian with additional terms describing the axion field and its coupling to photons. The complete action reads:
\begin{equation}
\mathcal{L}=R-\frac{1}{4}F_{\mu\nu}F^{\mu\nu}-A_\mu J_e^\mu+\mathcal{L}_\varphi+\mathcal{L}_{\text{int}},
\end{equation}
where $R$ is the Ricci scalar, $F_{\mu\nu}$ denotes the tensor of the electromagnetic field strength and $J_e^\mu$ represents the four-current electric. The axion field $\varphi$ enters through its kinetic and mass terms $\mathcal{L}_\varphi = \nabla_\mu \varphi^* \nabla^\mu \varphi - m_\varphi^2 |\varphi|^2$, while its interaction with the electromagnetic sector is encoded in the Chern-Simons-type coupling $\mathcal{L}_{\text{int}} = -\frac{g}{4} \varepsilon^{\mu\nu\alpha\beta} F_{\alpha\beta} F_{\mu\nu}$, with $g$ being the axion-photon coupling constant. This interaction term violates parity and leads to distinctive polarization-dependent effects on photon propagation.

In the presence of a magnetized plasma permeated by an axion background, photon trajectories deviate from null geodesics of the spacetime metric. The propagation is instead governed by an effective Hamiltonian that incorporates the refractive properties of the medium \cite{Synge:1960ueh}:
\begin{equation}
\mathcal{H}(x^\alpha, p_\alpha)=\frac{1}{2}\left[ g^{\alpha \beta} p_\alpha p_\beta - (n^2-1)( p_\beta u^\beta )^2 \right],
\end{equation}
where $n$ denotes the local refractive index, $x^\alpha$ are the spacetime coordinates, and $p_\alpha$, $u^\beta$ represent the four-momentum photon and four-velocity plasma, respectively. The frequency-dependent refractive index acquires contributions from both the plasma response and the axion-photon mixing, as derived in \cite{Mendonca:2019eke}:
\begin{eqnarray}
n^2&=&1- \frac{\omega_{\text{p}}^2}{\omega^2}-\frac{f_0}{\gamma_{0}}\frac{\omega_{\text{p}}^2}{(\omega-k u_0)^2}-\frac{\Omega^4}{\omega^2(\omega^2-\omega_{\varphi}^2)}\nonumber \\
&&-\frac{f_0}{\gamma_{0}}\frac{\Omega^4}{(\omega-k u_0)^2(\omega^2-\omega_{\varphi}^2)},
\label{eq:n1}
\end{eqnarray}
where $\omega_{\text{p}}^2=4\pi e^2 N/m_e$ is the plasma frequency determined by the density of the electron number $N$, and $\Omega=(gB_0\omega_p)^{1/2}$ characterizes the strength of the coupling of axion-photons in the presence of an external magnetic field $B_0$. The quantity $\omega_\varphi = m_\varphi c^2/\hbar$ represents the Compton frequency of the axion, which sets the energy scale for resonant conversion between axions and photons.

For analytical tractability, we consider a static plasma ($f_0 = 0$) and obtain the simplified expression:
\begin{eqnarray}
n^2(r)=1- \frac{\omega_{\text{p}}^2(r)}{\omega(r)^2} \left(1+\frac{g^2 B_0^2}{\omega(r)^2-\omega_{\varphi}^2}\right).
\end{eqnarray}
The term in parentheses reveals the essential physics: the standard plasma contribution $\omega_p^2/\omega^2$ is enhanced by a factor that depends on the axion parameters and diverges as the photon frequency approaches the axion mass ($\omega \to \omega_\varphi$). This resonance structure underlies the potential for dramatic lensing modifications in environments where the axion mass matches typical photon energies.

Accounting for gravitational redshift through the relation $\omega(r) = \omega_0/\sqrt{f(r)}$, where $\omega_0$ is the photon frequency at infinity and $f(r)$ is the metric function of the confining NED BH given in Eq.~\eqref{metricFun}, the refractive index becomes \cite{Atamurotov:2021cgh,sucu2024effect}:
\begin{equation}
n(r)\simeq\sqrt{ 1-\frac{\omega_{\text{p}}^2}{\omega_0^2}f(r) \left(1+\frac{ B^2_0 }{1-\omega_{\varphi}^2}\right)}.
\end{equation}
This expression enables the construction of an optical metric that captures the combined effects of gravity, plasma dispersion, and axion-photon mixing:
\begin{eqnarray}
dt^{2}=\left[1-\frac{\omega_{\text{p}}^2}{\omega_0^2}f(r) \left(1+\frac{ B^2_0 }{1-\omega_{\varphi}^2}\right)\right]\left[ \frac{dr^2}{f(r)^2} +\frac{r^2}{f(r)}d\phi^2\right].
\end{eqnarray}

The Gaussian curvature $\tilde{\mathcal{K}}$ associated with this optical geometry of the axion-plasma takes the form:
\begin{equation}
\tilde{\mathcal{K}}= \frac{\left(\omega_{\varphi}^{2}-1\right) M_{\text{ADM}} \omega_{0}^{2}}{\left(B_{0}^{2}-\omega_{\varphi}^{2}+1\right) \omega_{p}^{2} r^{3}}.
\end{equation}
Several features of this expression merit attention. The numerator contains the factor $(\omega_\varphi^2 - 1)$, which changes sign depending on whether the axion mass is above or below a characteristic scale set by the photon frequency. The denominator includes the combination $(B_0^2 - \omega_\varphi^2 + 1)$, which can approach zero under specific resonance conditions, potentially leading to large enhancements of the curvature and hence the deflection angle.

Applying the GBT to this optical geometry yields the total deflection angle in the axion-plasmon medium:
{\small
\begin{multline}
\Theta \simeq -\int_0^\pi \int_{b/\sin\phi}^\infty \tilde{\mathcal{K}} \, dS = \frac{2 M_{\text{ADM}}}{b \left(B_{0}^{2}-\omega_{\varphi}^{2}+1\right)}+\frac{2 M_{\text{ADM}} B_{0}^{2}}{b \left(B_{0}^{2}-\omega_{\varphi}^{2}+1\right)}\\
-\frac{2 M_{\text{ADM}} \omega_{0}^{2}}{b \left(B_{0}^{2}-\omega_{\varphi}^{2}+1\right) \omega_{p}^{2}}+\frac{2 M_{\text{ADM}} \omega_{0}^{2} \omega_{\varphi}^{2}}{b \left(B_{0}^{2}-\omega_{\varphi}^{2}+1\right) \omega_{p}^{2}}-\frac{2 M_{\text{ADM}} \omega_{\varphi}^{2}}{b \left(B_{0}^{2}-\omega_{\varphi}^{2}+1\right)}\\
+\frac{M_{\text{ADM}}^{2} \pi}{4 b^{2} \left(B_{0}^{2}-\omega_{\varphi}^{2}+1\right)}+\frac{M_{\text{ADM}}^{2} \pi  B_{0}^{2}}{4 b^{2} \left(B_{0}^{2}-\omega_{\varphi}^{2}+1\right)}\\
-\frac{3 M_{\text{ADM}}^{2} \pi  \omega_{0}^{2}}{4 b^{2} \left(B_{0}^{2}-\omega_{\varphi}^{2}+1\right) \omega_{p}^{2}}+\frac{3 M_{\text{ADM}}^{2} \pi  \omega_{0}^{2} \omega_{\varphi}^{2}}{4 b^{2} \left(B_{0}^{2}-\omega_{\varphi}^{2}+1\right) \omega_{p}^{2}}-\frac{M_{\text{ADM}}^{2} \pi  \omega_{\varphi}^{2}}{4 b^{2} \left(B_{0}^{2}-\omega_{\varphi}^{2}+1\right)}.
\end{multline}}

This result encapsulates the interplay among gravitational lensing, plasma dispersion, magnetic field effects, and axion-photon coupling in a single analytical expression. The deflection angle exhibits several notable properties that distinguish it from the pure vacuum or plasma cases analyzed in Sec.~\ref{isec3}. First, all terms carry a common denominator $(B_0^2 - \omega_\varphi^2 + 1)$ that introduces a resonance structure: when the magnetic field strength and axion mass satisfy $B_0^2 + 1 \approx \omega_\varphi^2$, the deflection can become arbitrarily large, signaling a breakdown of the weak-field approximation and the onset of strong axion-photon conversion. Second, the frequency dependence encoded in the ratios $\omega_0^2/\omega_p^2$ and $\omega_\varphi^2$ implies that observations at multiple wavelengths could be used to disentangle the various contributions and extract constraints on the axion parameters.

The presence of the axion field introduces an energy-dependent modulation of the deflection angle that differs qualitatively from standard plasma chromatic effects. While a pure plasma medium produces a $1/\omega_0^2$ frequency scaling, the axion contribution exhibits a more complex dependence through the resonance factor involving $\omega_\varphi$. Near the resonance condition $\omega \approx \omega_\varphi$, the deflection angle can be significantly enhanced or suppressed depending on the relative signs of the various terms, offering a potential observational signature of axionic DM in the vicinity of compact magnetized objects.

From an observational perspective, these results suggest that precision gravitational lensing measurements around strongly magnetized BHs—such as those powering active galactic nuclei or residing in X-ray binary systems—could serve as probes of beyond-Standard Model physics. The confining NED BH considered in this work provides a particularly interesting testbed because the NED corrections to the metric function (scaling as $Q^3/\xi$) introduce additional parameter dependence that could help break down the degeneracies between gravitational and medium effects. Multi-frequency observations spanning radio to X-ray bands, combined with polarimetric data sensitive to axion-induced birefringence, would maximize the constraining power of such measurements \cite{o2024cosmology,sushkov2023quantum}.

\section{Gravitational Redshift of the Confining NED BH} \label{isec5}

Gravitational redshift stands as one of the earliest and most direct tests of GR, providing a clean observational probe of spacetime curvature near massive objects \cite{pound1960apparent,will2014confrontation}. When photons escape from a gravitational potential well, they lose energy and their frequency decreases—an effect first measured in the classic Pound-Rebka experiment and now routinely observed in signals from neutron stars, white dwarfs, and the vicinity of BHs \cite{pound1959gravitational}. For the confining NED BH studied in this work, the $r^{-4}$ correction to the metric function introduces a characteristic modification to the redshift formula that distinguishes this geometry from both Schwarzschild and standard RN spacetimes.

Consider a static, spherically symmetric line element of the form given in Eq.~\eqref{metric}. The Killing vector $\partial_t$ generates a conserved quantity along the geodesics, which for photons corresponds to the energy $\mathcal{E}$ measured at infinity. A static observer located at radius $r$ measures the photon frequency as \cite{wald2024general}:
\begin{equation}\label{eq:omega_static_NED}
\omega(r)=\frac{\mathcal{E}}{\sqrt{f(r)}},
\end{equation}
where $f(r)$ is the metric function defined in Eq.~\eqref{metricFun}. This expression shows that the locally measured frequency increases as $f(r)$ decreases—that is, deeper in the gravitational potential, the same photon appears to be more energetic to a local observer.

The redshift $z$ experienced by a photon emitted at radius $r_e$ and received by an observer at radius $r_o$ follows directly from comparing the frequencies at these two locations:
\begin{equation}\label{eq:general_z_NED}
1+z=\frac{\omega(r_e)}{\omega(r_o)}=\sqrt{\frac{f(r_o)}{f(r_e)}}.
\end{equation}
For an observer situated far from the BH, where the spacetime asymptotically approaches Minkowski geometry and $f(r_o) \to 1$, the redshift simplifies to:
\begin{equation}\label{eq:z_infty_exact_NED}
z_{\infty}=\frac{1}{\sqrt{f(r_e)}}-1.
\end{equation}
This compact formula encodes all the information about how the confining NED geometry affects the frequency of escaping radiation. The entire effect is controlled by the metric function evaluated at the emission point $r_e$.

To extract the separate contributions from the gravitational mass and the NED electromagnetic structure, we define the deviation from flat spacetime:
\begin{equation}
\varepsilon(r)=f(r)-1=-\frac{2M_{\mathrm{ADM}}}{r}+\frac{Q^{3}}{9\xi^{2}r^{4}}.
\end{equation}
The first term represents the standard Newtonian potential (with relativistic corrections absorbed into $M_{\text{ADM}}$), while the second term captures the NED contribution that decays as $r^{-4}$. Unlike the RN metric, where the electromagnetic correction appears as $Q^2/r^2$, the confining NED model produces a much faster falloff, making the correction negligible at large distances, but potentially significant near the EH.

In the weak-field regime where $|\varepsilon| \ll 1$, a Taylor expansion of the square root yields:
\begin{equation}
\frac{1}{\sqrt{1+\varepsilon}}=1-\frac{\varepsilon}{2}+\frac{3}{8}\varepsilon^{2}+\mathcal{O}(\varepsilon^{3}),
\end{equation}
and substituting this into Eq.~\eqref{eq:z_infty_exact_NED} gives:
\begin{equation}\label{eq:z_expanded_NED}
z_{\infty}\approx -\frac{\varepsilon(r_e)}{2}+\frac{3}{8}\varepsilon(r_e)^{2}.
\end{equation}
Keeping only the leading-order terms and separating the mass and charge contributions, we obtain:
\begin{equation}\label{eq:z_linear_NED}
z_{\infty}\approx \frac{M_{\mathrm{ADM}}}{r_e}-\frac{Q^{3}}{18\,\xi^{2} r_e^{4}}
+\mathcal{O}\!\left(\frac{M_{\mathrm{ADM}}^{2}}{r_e^{2}},
\frac{M_{\mathrm{ADM}}Q^{3}}{\xi^{2}r_e^{5}},
\frac{Q^{6}}{\xi^{4}r_e^{8}}\right).
\end{equation}

This expression admits a clear physical interpretation. The first term, $M_{\text{ADM}}/r_e$, reproduces the classical Schwarzschild redshift that scales inversely with the emission radius. The second term, proportional to $-Q^3/(\xi^2 r_e^4)$, represents the NED correction and carries a negative sign. This means that the confining electromagnetic field partially counteracts the gravitational redshift, reducing the total frequency shift compared to a Schwarzschild BH of the same ADM mass. The effect becomes more pronounced at smaller emission radii where the $r^{-4}$ scaling dominates over the $r^{-1}$ term.

The magnitude of the NED correction depends on the ratio $Q^3/\xi^2$. For fixed magnetic charge $Q$, increasing the NED parameter $\xi$ suppresses the correction, pushing the redshift closer to the Schwarzschild value. Conversely, for fixed $\xi$, larger values of $Q$ enhance the NED contribution. This parameter dependence mirrors what we found in the horizon structure (Table~\ref{tab:NED_horizons}) and the analysis of the deflection angle (Sect.~\ref{isec3}), confirming that $Q$ and $\xi$ act as competing influences on the observable properties of the confining NED BH.

Figure~\ref{red} displays the gravitational redshift $z_\infty$ as a function of the emission radius $r_e$ and the NED parameter $\xi$, with the Schwarzschild mass and magnetic charge fixed at $M=1$ and $Q=0.5$, respectively. The surface plot exhibits the expected behavior: the redshift increases sharply as the emission point approaches the EH (small $r_e$) and decreases monotonically with increasing distance from the BH. For fixed $r_e$, the variation with $\xi$ reflects two competing effects. On the one hand, larger $\xi$ increases the ADM mass through the logarithmic correction in Eq.~(7), which tends to enhance the redshift. On the other hand, a larger $\xi$ also suppresses the $Q^3/\xi^2$ correction term, which reduces the contribution of the NED to the redshift. The net effect depends on which mechanism dominates at a given radius, leading to the nontrivial surface structure visible in Fig.~\ref{red}.

\begin{figure}[t]
    \centering
    \includegraphics[width=0.5\textwidth]{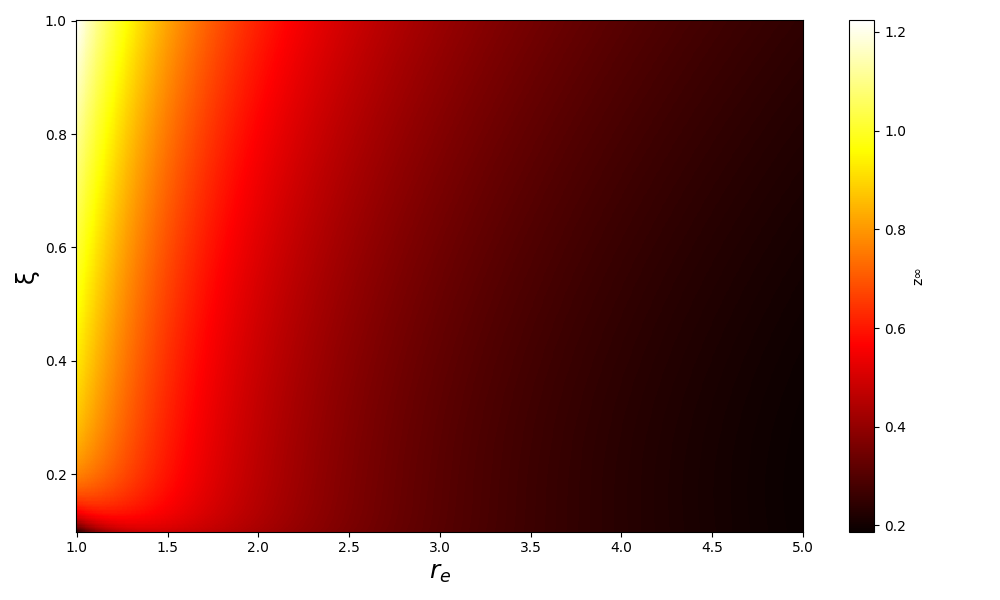}
    \caption{Gravitational redshift $z_\infty$ as a function of the emission radius $r_e$ and the NED parameter $\xi$, for fixed values $M=1$ and $Q=0.5$. The redshift increases steeply near the EH and decreases at large $r_e$, recovering the weak-field limit. The variation with $\xi$ reflects the competition between the enhanced ADM mass and the suppressed NED correction term.}
    \label{red}
\end{figure}

From an observational standpoint, the $r^{-4}$ dependence of the NED correction offers a potential method to distinguish this BH model from standard GR predictions. In Schwarzschild spacetime, the redshift depends only on the mass-to-radius ratio $M/r_e$, while in RN geometry an additional $Q^2/r_e^2$ term appears. The confining NED BH, by contrast, introduces a correction that falls off much more rapidly with distance. Measurements of redshifted emission lines from matter orbiting close to the EH—such as iron K$\alpha$ lines observed by X-ray satellites—could in principle detect this steeper radial dependence, provided sufficient spectral resolution and signal-to-noise ratio are available \cite{fabian1989x,reynolds2003fluorescent}.

The redshift analysis also connects to the thermodynamic properties of the BH. The Hawking temperature, derived from surface gravity in the EH, depends on the derivative $f'(r_h)$ evaluated at the horizon. Since the NED correction modifies both $f(r)$ and its derivative, the thermal emission spectrum will carry the imprint of the confining electromagnetic structure. We explore these thermodynamic aspects in detail in the following section.

In short, gravitational redshift provides a direct window into the metric function of the confining NED BH. The characteristic $r^{-4}$ correction distinguishes this geometry from both Schwarzschild and RN spacetimes, producing a reduced redshift at fixed ADM mass due to the negative sign of the NED contribution. The parameter dependence on $Q$ and $\xi$ mirrors found in previous sections reinforces the picture of a BH geometry shaped by the competition between gravitational attraction and the confining electromagnetic field.

\section{EC JTE and Heat Capacity of the Confining NED BH} \label{isec6}

BH thermodynamics has evolved considerably since the pioneering work of Bekenstein and Hawking, who established that BHs possess an entropy proportional to their horizon area and emit thermal radiation at a characteristic temperature \cite{bekenstein1973black,hawking1974black}. In the extended phase space formalism, the cosmological constant $\Lambda$ is promoted to a thermodynamic variable identified with pressure $P = -\Lambda/(8\pi)$, while the BH mass plays the role of enthalpy rather than internal energy \cite{kruglov2022nonlinearly,kruglov2022ned,kruglov2023magnetic,dolan2011cosmological}. This framework enables the study of processes such as the JTE, which describes how a system's temperature changes during isenthalpic (constant enthalpy) throttling. For ordinary gasses, the JTE determines whether expansion causes cooling or heating; for BHs, the analogous process reveals information about the phase structure and thermal response of the gravitational system \cite{silva2021joule,yasir2023thermal}.

The confining NED BH derived in Sec.~\ref{isec2} provides an interesting arena for thermodynamic analysis because the $r^{-4}$ correction to the metric function directly modifies the surface gravity and hence the Hawking temperature. Before examining the JTE, we first establish the basic thermodynamic quantities. The Hawking temperature follows from the surface gravity $\kappa$ in the outer EH, located at $r = r_h$:
\begin{equation}
T_H = \frac{\kappa}{2\pi} = \frac{f'(r_h)}{4\pi}.
\end{equation}
For the metric function given in Eq.~\eqref{metricFun}, the derivative is evaluated to:
\begin{equation}
f'(r) = \frac{2M_{\text{ADM}}}{r^2} - \frac{4Q^3}{9\xi^2 r^5},
\end{equation}
so that the Hawking temperature at the horizon becomes:
\begin{equation}
T_H = \frac{1}{4\pi} \left( \frac{2M_{\text{ADM}}}{r_h^2} - \frac{4Q^3}{9\xi^2 r_h^5} \right).
\end{equation}
The first term reproduces the Schwarzschild result $T_H = 1/(8\pi M)$ when $Q \to 0$, while the second term represents the NED correction. Since this correction carries a negative sign, the confining electromagnetic field tends to reduce the Hawking temperature compared to a Schwarzschild BH of equal ADM mass, provided the horizon radius remains fixed.

The Bekenstein-Hawking entropy takes the standard area law form:
\begin{equation}
S = \frac{A}{4} = \pi r_h^2,
\end{equation}
where $A = 4\pi r_h^2$ is the area of the horizon and {the exponantial entropy is of the form \cite{sucu2025exploring}
\begin{equation}
    S_{EC}=S+e^S
\end{equation}
In the extended phase space, the thermodynamic volume conjugate to pressure is given by:
\begin{equation}
V = \frac{4}{3}\pi r_h^3.
\end{equation}
The JTE describes how temperature varies with pressure during a process in which enthalpy $H$ (identified with the BH mass $M$) remains constant. The central quantity characterizing this process is the Joule-Thomson coefficient \cite{liang2021joule,kruglov2022nonlinearly,aydiner2025regular}:
\begin{equation}
\mu_J = \left( \frac{\partial T_H}{\partial P} \right)_H,
\end{equation}
where the subscript $H$ indicates that the derivative is taken at fixed enthalpy. The sign of $\mu_J$ determines the thermal response: when $\mu_J > 0$, the BH cools upon expansion (pressure decrease), analogous to the behavior of most real gasses below their inversion temperature. When $\mu_J < 0$, expansion leads to heating instead. The boundary $\mu_J = 0$ defines the inversion curve in the $T$-$P$ plane, separating the cooling and heating regimes.

Using standard thermodynamic identities, the JT coefficient can be rewritten as \cite{sucu2025quantumOzcan}:
\begin{equation}
\mu_J = \frac{1}{C_P} \left[ T_H \left( \frac{\partial V}{\partial T_H} \right)_P - V \right],
\end{equation}
where $C_P$ denotes the heat capacity at constant pressure. For BH systems where the thermodynamic quantities are naturally expressed as functions of the horizon radius $r_h$, it proves convenient to use the chain rule:
\begin{equation}
\mu_J = \left( \frac{d T_H}{d r_h} \right) \left( \frac{d P_{EC}}{d r_h} \right)^{-1}.
\end{equation}
This formulation bypasses the need to invert $T_H(r_h)$ and directly yields $\mu_J$ from the known functional forms.

Carrying out the calculation for the confining NED BH, we obtain:
\begin{equation}
    \mu_J \approx -\frac{72 \pi  \left(M_{\text{ADM}} \xi  r_h^{3}-\frac{5 Q^{3}}{9}\right) r_h^{3}}{9 M_{\text{ADM}} \pi^{2} r_h^{7} \xi -20 \pi^{2} Q^{3} r_h^{4}+90 M_{\text{ADM}} \xi  r_h^{3}-80 Q^{3}}.
\end{equation}
The numerator vanishes when $M_{\text{ADM}} \xi r_h^3 = 5Q^3/9$, which defines the inversion point where the BH transitions from cooling to heating behavior (or vice versa). The denominator determines the divergence structure and is related to the heat capacity, as we discuss below.

Figure~\ref{jte} displays the JT coefficient $\mu_J$ as a function of the radius of the horizon $r_h$ and the NED parameter $\xi$, with $M=1$ and $Q=0.5$. The surface exhibits positive and negative regions, separated by the inversion curve where $\mu_J = 0$. For small horizon radii (corresponding to hot, small BHs), the coefficient tends to be negative, indicating that these BHs heat up upon expansion. For larger horizons, $\mu_J$ becomes positive and the BH cools when the pressure decreases. The location of the inversion curve changes with $\xi$: Higher values of the NED parameter push the transition to smaller horizon radii, reflecting the enhanced ADM mass contribution that modifies the thermal response.
\begin{figure}[t]
    \centering
    \includegraphics[width=0.5\textwidth]{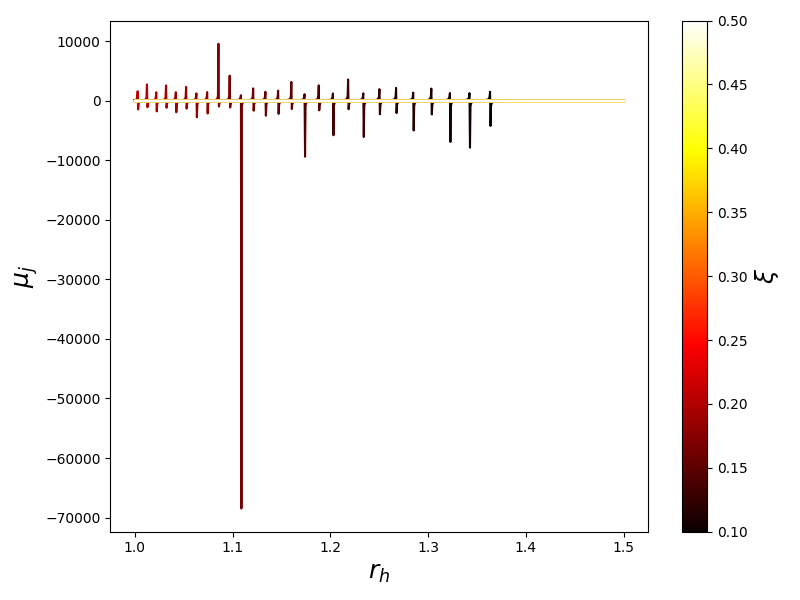}
    \caption{JT coefficient $\mu_J$ as a function of the horizon radius $r_h$ and the NED parameter $\xi$, for fixed values $M=1$ and $Q=0.5$. Positive values indicate cooling upon expansion, while negative values correspond to heating. The zero-crossing defines the inversion curve separating these two regimes.}
    \label{jte}
\end{figure}

The heat capacity provides complementary information about the thermal stability. A BH with positive heat capacity can reach thermal equilibrium with a heat bath: if it loses energy through Hawking radiation, its temperature decreases, reducing further emission and allowing equilibration. Conversely, a BH with negative heat capacity exhibits runaway behavior—energy loss increases the temperature, accelerating the evaporation process \cite{synge1966escape}. The heat capacity at constant pressure is defined as:
\begin{equation}
C_P = T_{H}\left(\frac{\partial S}{\partial T_{H}}\right)_P. \label{s25}
\end{equation}

For the confining NED BH, evaluating this expression yields the following:
\begin{equation}
    C_P \approx -\frac{\pi^{2} \left(9 M_{\text{ADM}} \xi  r_h^{3}-2 Q^{3}\right) r_h^{4}}{9 M_{\text{ADM}} \xi  r_h^{3}-5 Q^{3}}.
\end{equation}
The sign of $C_P$ is determined by the ratio of the numerator and denominator. The numerator vanishes when $9 M_{\text{ADM}} \xi r_h^3 = 2Q^3$, while the denominator vanishes when $9 M_{\text{ADM}} \xi r_h^3 = 5Q^3$. Between these two critical radii, the heat capacity changes sign, indicating a phase transition from thermally stable to unstable configurations (or vice versa).

Figure~\ref{cc} shows the heat capacity $C_P$ as a function of $r_h$ and $\xi$ for the same parameter values $M=1$ and $Q=0.5$. The surface displays a characteristic divergence at the critical radius where the denominator vanishes, corresponding to a second-order phase transition in the canonical ensemble. On one side of this divergence, $C_P > 0$ and the BH is thermally stable; on the other side, $C_P < 0$ and the BH becomes unstable against thermal fluctuations. The position of the divergence depends on $\xi$: increasing the NED parameter shifts the critical point, altering the size of the stable and unstable branches.

\begin{figure}[t]
    \centering
    \includegraphics[width=0.5\textwidth]{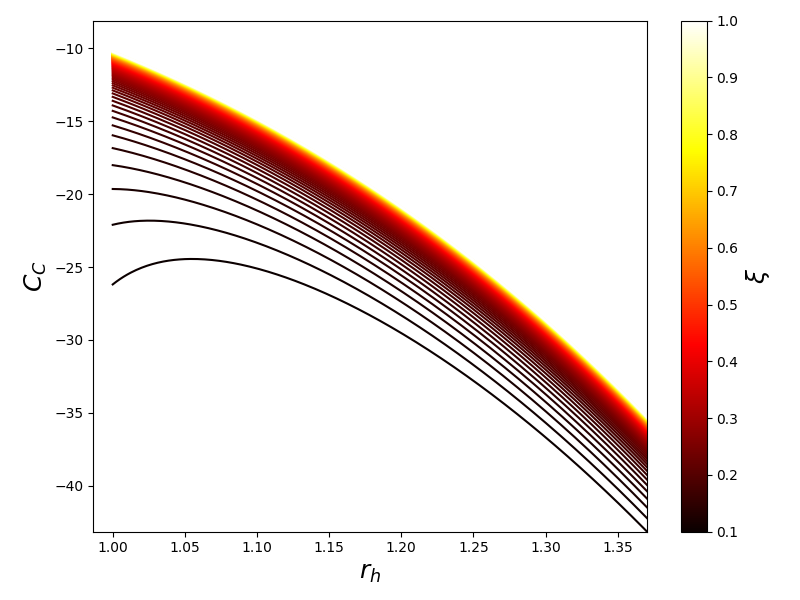}
    \caption{Heat capacity $C_P$ as a function of the horizon radius $r_h$ and the NED parameter $\xi$, for fixed values $M=1$ and $Q=0.5$. The divergence indicates a phase transition between thermally stable ($C_P > 0$) and unstable ($C_P < 0$) configurations. The location of the critical point shifts with $\xi$.}
    \label{cc}
\end{figure}

The connection between the JT coefficient and the heat capacity is evident by comparing Figs.~\ref{jte} and \ref{cc}. The divergence in $C_P$ corresponds to a sign change in $\mu_J$, as the heat capacity appears in the denominator of the standard JT formula. This correlation reflects the underlying thermodynamic structure: phase transitions marked by divergent heat capacity necessarily affect the isenthalpic expansion behavior. In general, the JTE and heat capacity analysis reveals that the confining NED BH exhibits nontrivial thermal properties shaped by the competition between the ADM mass term and the $Q^3/\xi^2$ correction. The inversion temperature that separates the cooling and heating regimes, and the critical point that marks the onset of thermal instability, both depend on the NED parameters in ways that distinguish this model from Schwarzschild or RN BHs. These thermodynamic signatures, combined with the lensing and redshift results from previous sections, provide multiple independent probes of the confining electromagnetic structure.

\section{Lyapunov Exponent for Circular Geodesics} \label{isec7}

The stability of particle orbits around BHs provides valuable information on the underlying spacetime geometry and has direct observational consequences for phenomena such as BH shadows, photon rings, and QNM spectra \cite{johnson2020universal,event2019first,sucu2025quantumHassan}. Circular null geodesics, which form the photon sphere, are particularly important because they define the boundary of the BH shadow and control the eikonal limit of QNM frequencies \cite{bozza2002gravitational,falcke1999viewing}. In this section, we analyze the stability of circular orbits in the confining NED BH spacetime and derive the Lyapunov exponent that characterizes the rate at which nearby trajectories diverge from the unstable photon orbit.

Consider the line element given in Eq.~\eqref{metric} with the metric function defined in Eq.~\eqref{metricFun}. Restricting attention to equatorial motion ($\theta = \pi/2$), the geodesic Lagrangian takes the form:
\begin{equation}
    \mathcal{L}=\frac{1}{2}\left[-f(r)\dot{t}^{2}+\frac{\dot{r}^{2}}{f(r)}+r^{2}\dot{\varphi}^{2}\right],
\end{equation}
where overdots denote derivatives with respect to an affine parameter. The metric admits two Killing vectors, $\partial_t$ and $\partial_\varphi$, corresponding to time translation and axial symmetry. These symmetries generate conserved quantities along geodesics:
\begin{equation}
    E=f(r)\dot{t}, \qquad L=r^{2}\dot{\varphi},
\end{equation}
where $E$ represents the specific energy and $L$ the specific angular momentum of the orbiting particle.

Substituting these conserved quantities into the normalization condition $g_{\mu\nu}\dot{x}^\mu\dot{x}^\nu = -\varepsilon$ yields the radial equation of motion in the form:
\begin{equation}
    \dot{r}^{2}+V_{\rm eff}(r)=E^{2},
\end{equation}
where the effective potential governing radial motion is:
\begin{equation}
    V_{\rm eff}(r)=f(r)\left(\frac{L^{2}}{r^{2}}-\varepsilon\right).
\end{equation}
Here $\varepsilon = 0$ for null geodesics (photons) and $\varepsilon = -1$ for timelike geodesics (massive particles). The effective potential encodes how the NED correction to the metric function influences orbital dynamics through the factor $f(r)$.

Circular orbits occur at radii $r_0$ where the effective potential satisfies:
\begin{equation}
    V_{\rm eff}'(r_{0})=0, \qquad V_{\rm eff}''(r_{0})<0,
\end{equation}
with the first condition ensuring that the radial velocity vanishes and the second selecting unstable orbits where small perturbations grow exponentially. For photons ($\varepsilon = 0$), the condition $V_{\rm eff}'(r_0) = 0$ leads to:
\begin{equation}
    \frac{d}{dr}\left[\frac{f(r)}{r^2}\right]_{r=r_0} = 0,
\end{equation}
which, upon substituting the confining NED metric function, yields the quartic equation:
\begin{equation}
    r_{0}^{4}-3M_{\rm ADM}r_{0}^{3}+\frac{Q^{3}}{3\xi^{2}}=0.
    \label{eq:photon_sphere_equation}
\end{equation}
This equation determines the photon sphere radius $r_0$—the location where light can orbit the BH on unstable circular paths. The structure of Eq.~\eqref{eq:photon_sphere_equation} differs markedly from both Schwarzschild geometry, where $r_{\rm ph} = 3M$, and RN spacetime, where the photon sphere satisfies a cubic equation. The quartic nature arises from the $r^{-4}$ correction in the confining NED metric and introduces additional parameter dependence through the combination $Q^3/\xi^2$.

The physical solutions of Eq.~\eqref{eq:photon_sphere_equation} must satisfy $r_0 > r_h$ (the photon sphere lies outside the EH) and yield real, positive values. For small NED corrections, the photon sphere radius can be expanded perturbatively around the Schwarzschild value:
\begin{equation}
r_0 \approx 3M_{\rm ADM} - \frac{Q^3}{27 M_{\rm ADM}^2 \xi^2} + \mathcal{O}\left(\frac{Q^6}{\xi^4}\right).
\end{equation}
The negative sign of the correction indicates that the confining electromagnetic field pulls the photon sphere inward compared to a Schwarzschild BH of equal ADM mass.

The instability of the photon orbit is quantified by the Lyapunov exponent $\lambda$, which measures the exponential growth rate of small radial perturbations \cite{event2019first,akiyama2019first}. Linearizing the radial equation around the circular orbit by writing $r = r_0 + \delta r$ with $|\delta r| \ll r_0$ gives the following:
\begin{equation}
    \ddot{\delta r}+\frac{1}{2}V_{\rm eff}''(r_{0})\,\delta r=0.
\end{equation}
Since $V_{\rm eff}''(r_0) < 0$ for unstable orbits, the solution grows exponentially as $\delta r \propto e^{\lambda \tau}$, where $\tau$ is the affine parameter. Converting to coordinate time $t$ using the relation $\dot{t} = L/(r_0\sqrt{f(r_0)})$ valid for null geodesics, the Lyapunov exponent becomes:
\begin{equation}
    \lambda=\sqrt{-\frac{r_{0}^{2}f(r_{0})}{2L^{2}}\,V_{\rm eff}''(r_{0})}.
\end{equation}

Evaluating $V_{\rm eff}''(r)$ explicitly for the confining NED metric and imposing the photon sphere condition Eq.~\eqref{eq:photon_sphere_equation} leads to a closed-form expression:
\begin{equation}
    \lambda^{2}=
    \frac{\big(r_{0}^{4}\xi^{2}-Q^{3}\big)\,\big(3r_{0}^{4}\xi^{2}-Q^{3}\big)}
         {9r_{0}^{10}\xi^{4}},
\end{equation}
which can be written as:
\begin{equation}
    \lambda=
    \frac{1}{3r_{0}^{5}\xi^{2}}
    \sqrt{\big(r_{0}^{4}\xi^{2}-Q^{3}\big)\big(3r_{0}^{4}\xi^{2}-Q^{3}\big)}.
\end{equation}
For the Lyapunov exponent to be real and positive, the argument of the square root must be non-negative. This requires $r_0^4 \xi^2 > Q^3$ and $3r_0^4 \xi^2 > Q^3$ (both positive factors) or both negative factors. The first case corresponds to physically relevant BH configurations where the photon sphere lies well outside the region dominated by NED corrections.

The Lyapunov exponent has direct observational implications. In the eikonal approximation, the real part of BH QNM frequencies is related to the angular velocity at the photon sphere, while the imaginary part (decay rate) is determined by the Lyapunov exponent:
\begin{equation}
\omega_{\rm QNM} \approx \Omega_0 \ell - i\left(n + \frac{1}{2}\right)\lambda,
\end{equation}
where $\Omega_0 = \sqrt{f(r_0)}/r_0$ is the orbital frequency, $\ell$ is the quantum number of angular momentum, and $n$ is the overtone number. Thus, measurements of QNM ringdown signals from BH mergers could in principle constrain the NED parameters $Q$ and $\xi$ through their effect on $\lambda$.

The expression for $\lambda$ reveals how the confining geometry of the NED differs from standard BH solutions. In Schwarzschild spacetime, $\lambda_{\rm Sch} = 1/(3\sqrt{3}M)$, while for RN BHs the Lyapunov exponent depends on both $M$ and $Q$ through a different functional form. The confining NED BH introduces the additional parameter $\xi$, and the dependence on $Q^3/\xi^2$ rather than $Q^2$ produces a distinct scaling behavior. For fixed $Q$, increasing $\xi$ reduces the NED correction and pushes $\lambda$ toward the Schwarzschild limit. Conversely, for fixed $\xi$, increasing $Q$ improves the departure from GR predictions.

Therefore, one can deduce that the analysis of circular geodesics in the confining NED spacetime shows that the $Q^3/(9\xi^2 r^4)$ term in the metric function modifies both the photon sphere location and the Lyapunov exponent characterizing orbital instability. The quartic structure of the photon sphere equation and the explicit $\xi$-dependence of $\lambda$ distinguish this geometry from Schwarzschild and RN BHs, offering potential observational signatures through QNM spectroscopy and photon ring measurements.

\section{Photon Sphere and Shadow of the Confining NED BH} \label{isec8}

The direct imaging of BH shadows by the EHT collaboration has opened a new window for testing gravity in the strong-field regime \cite{akiyama2019first,akiyama2022first}. The shadow—a dark silhouette surrounded by a bright photon ring—arises because light rays passing close to the BH are either captured or deflected away from the observer, leaving a characteristic deficit in the observed intensity. The size and shape of this shadow depend directly on the radius of the photon sphere and the underlying metric function, making shadow observations a promising tool for discriminating between different BH models \cite{akiyama2019first}. In this section, we compute the photon sphere location and shadow radius for the confining NED BH, based on the geodesic analysis presented in Sec.~\ref{isec7}.

The spacetime under consideration is described by the static, spherically symmetric line element:
\begin{equation}
ds^2 = -f(r)\, dt^2 + \frac{dr^2}{f(r)} + r^2 (d\theta^2 + \sin^2\theta\, d\phi^2),
\end{equation}
with the metric function:
\begin{equation}
f(r;\xi) = 1 - \frac{2 M_{\rm ADM}(\xi)}{r} + \frac{Q^3}{9 \xi^2 r^4},
\end{equation}
where $Q$ denotes the magnetic charge and $\xi$ is the NED parameter that controls the strength of the confining electromagnetic corrections. The ADM mass receives a contribution from the nonlinear electromagnetic self-energy:
\begin{equation}
M_{\rm ADM}(\xi) = M + \frac{2 \sqrt{2}}{3} \, \xi\, Q^{3/2} \ln \big(2 \sqrt{2} \xi Q^{1/2}\big).
\end{equation}

As derived in Sec.~\ref{isec7}, the photon sphere corresponds to unstable circular null geodesics where the effective potential for radial motion has a local maximum. The radius of the photon sphere $r_{\rm ph}$ satisfies the condition:
\begin{equation}
\left. \frac{d}{dr} \left( \frac{r^2}{f(r)} \right) \right|_{r=r_{\rm ph}} = 0,
\end{equation}
which can be rewritten as:
\begin{equation}
r_{\rm ph} f'(r_{\rm ph}) - 2 f(r_{\rm ph}) = 0.
\end{equation}
This relation follows from the requirement that the critical impact parameter $b_c = L/E$ corresponds to a turning point where incoming photons asymptotically spiral onto the photon sphere. For the confining NED metric, substituting the explicit form of $f(r)$ yields the quartic equation derived in Eq.~\eqref{eq:photon_sphere_equation}. Due to the transcendental dependence of $M_{\rm ADM}$ on $\xi$ through the logarithmic term, closed-form solutions are not available, and the radius of the photon sphere must be determined numerically.

We employ a bracketing method combined with Brent's root-finding algorithm to solve for $r_{\rm ph}$. This approach guarantees convergence and provides accurate results even when the metric function exhibits strong nonlinearity near the horizon. Starting from an initial bracket that contains the expected root (between the outer EH and the asymptotic region), the algorithm iteratively refines the estimate until the residual falls below a specified tolerance.

Once $r_{\rm ph}$ is determined, the shadow radius as perceived by a distant observer follows from geometric optics. Photons approaching the BH with impact parameter $b < b_c$ fall into the horizon, while those with $b > b_c$ are scattered to infinity. The critical impact parameter, which defines the apparent shadow boundary, equals:
\begin{equation}
b_{\rm ph} = \frac{r_{\rm ph}}{\sqrt{f(r_{\rm ph})}}.
\end{equation}
For a Schwarzschild BH, this formula gives $b_{\rm ph} = 3\sqrt{3}M \approx 5.196M$, which corresponds to a shadow angular diameter of about $10.4M/D$ where $D$ is the distance from the observer. The confining NED corrections modify both $r_{\rm ph}$ and $f(r_{\rm ph})$, leading to parameter-dependent deviations from this standard result.

Figure~\ref{fig:shadow_fullpage} presents synthetic shadow images for the confining NED BH at three different values of the coupling parameter: $\xi = 0.1$, $0.2$, and $0.3$, with fixed $M = 1$ and $Q = 0.5$. The images are constructed by ray-tracing null geodesics backward from the observer's screen and determining which rays terminate at infinity (bright) versus those captured by the BH (dark). The bright ring surrounding the shadow corresponds to photons that execute multiple orbits near the photon sphere before escaping—these form the so-called photon ring that carries information about the strong-field geometry \cite{johnson2020universal,akiyama2019first}.

\begin{figure*}[htbp]
    \centering
    \includegraphics[width=1.0\textwidth]{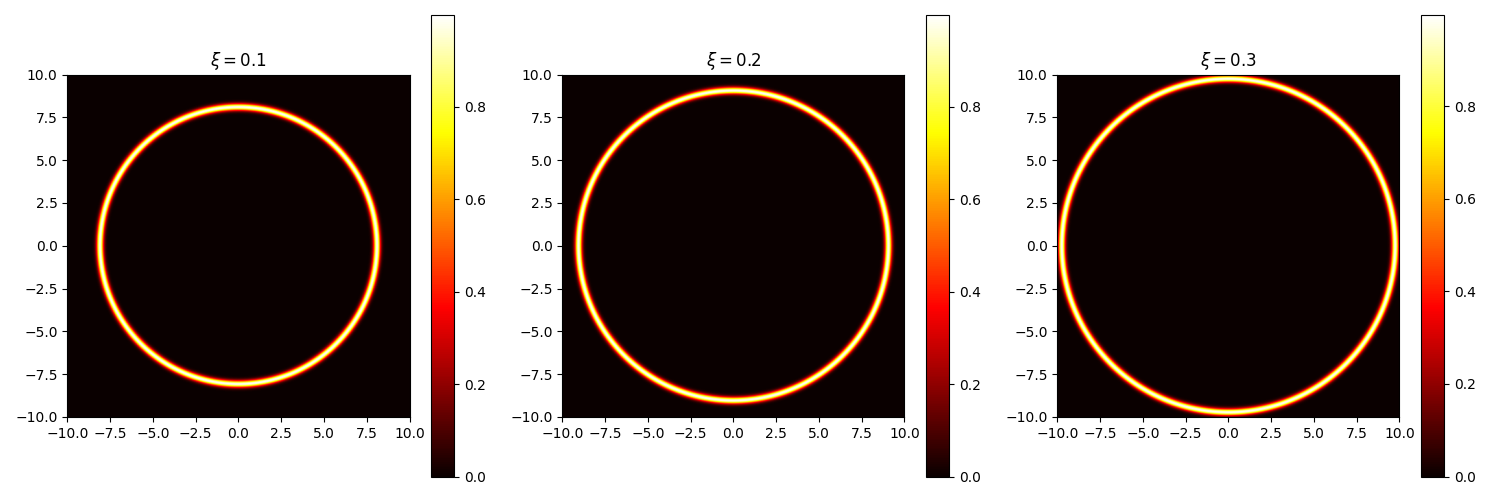}
    \caption{Synthetic shadow images of the confining NED BH for different values of the coupling parameter $\xi = 0.1$ (left), $\xi = 0.2$ (center), and $\xi = 0.3$ (right), with fixed $M = 1$ and $Q = 0.5$. The dark central region represents the shadow, while the bright Gaussian ring models the photon capture region formed by light rays that graze the photon sphere. Increasing $\xi$ enlarges the ADM mass and consequently expands both the photon sphere radius and the shadow diameter.}
    \label{fig:shadow_fullpage}
\end{figure*}

Several trends are visible in Fig.~\ref{fig:shadow_fullpage}. First, the shadow size increases monotonically with $\xi$. This behavior traces back to the ADM mass formula: larger $\xi$ enhances the logarithmic correction term, increasing $M_{\rm ADM}$ and thereby expanding both the EH and the photon sphere. Since the shadow radius scales roughly as $b_{\rm ph} \propto M_{\rm ADM}$, the shadow grows accordingly. Second, the shadow remains circular for all parameter values shown, as expected for a non-rotating, spherically symmetric spacetime. Deviations from circularity would require either spin (Kerr-like geometry) or non-spherical matter distributions \cite{gralla2019black,gralla2020measuring}.

Quantitatively, the shadow radius can be compared with the EHT measurements. For the supermassive BH M87*, the observed shadow diameter is approximately $42 \pm 3$ microarcseconds, corresponding to $(11 \pm 1.5)M$ when converted to gravitational units using the estimated mass $M \approx 6.5 \times 10^9 M_\odot$ \cite{johnson2020universal}. This value is slightly higher than the Schwarzschild prediction of $10.4M$, leaving room for deviations due to spin, environmental effects, or modifications to GR. The confining NED BH can accommodate such deviations through appropriate choices of $Q$ and $\xi$: increasing either parameter tends to enlarge the shadow, potentially improving agreement with observations.

The connection between the shadow and the Lyapunov exponent discussed in Sec.~\ref{isec7} deserves emphasis. The width of the photon ring—the bright feature encircling the shadow—is controlled by the Lyapunov exponent $\lambda$. Photons that approach the photon sphere from slightly different angles acquire exponentially growing separations in their orbital phases, leading to a demagnification factor $e^{-2\pi/\lambda}$ for successive photon ring images \cite{johnson2020universal,cardoso2009geodesic}. Thus, a larger Lyapunov exponent produces a narrower, more tightly wound photon ring structure. Since $\lambda$ depends on the NED parameters through Eq.~(53), measurements of photon ring properties could in principle constrain $Q$ and $\xi$ independently of the shadow size.

From an observational perspective, next-generation instruments such as the next-generation EHT (ngEHT) and space-based VLBI missions aim to resolve the photon ring substructure with sufficient precision to test strong-field gravity predictions \cite{johnson2023key}. The confining NED BH offers a concrete example of how NED modifications alter both the shadow boundary and the photon ring profile in ways that differ from Schwarzschild and RN geometries. The absence of the standard $Q^2/r^2$ term and its replacement by a $Q^3/(\xi^2 r^4)$ correction produces a distinct parameter scaling that could be distinguished observationally given sufficient angular resolution.\\

\section{Observable Intensities of Confining NED BH with Static Spherical Accretion}\label{isec8a}

In this section, we investigate the shadow structure and observable intensity profiles associated with confining NED BHs. For this purpose, we adopt the static spherical accretion model, which helps to produce a steady inflow of matter onto a confining NED BHs under the assumption of spherical symmetry and negligible angular momentum of the accreting gas. This section is particularly well suited to exploring confining NED BH shadows and intensity distributions, as it significantly simplifies the emission characteristics and enables an efficient evaluation of radiative properties.
By reducing the problem to a spherically symmetric configuration, the model allows for a transparent treatment of photon trajectories and radiative transfer, thereby facilitating the calculation of observable intensities. The absence of disk-related dynamics further helps to isolate the purely gravitational influence of the confining NED BHs on the emitted radiation, offering clearer insight into the underlying spacetime geometry. Assuming isotropic emission from the accreting matter in the static frame, the model effectively captures the roles of gravitational redshift and light bending in shaping the observed shadow. 

\subsection{Radiation Intensity and Emissivity} 
The observed specific intensities can be calculated by integrating the emissivity within the scope of the light ray trajectory \cite{Bambi:2013nla}, which is further expressed as:
\begin{equation}\label{41} I{(U^{s}_{\text{ob}})} =
\int_\gamma g^{s3} j(U_\text{em}^{s}) dl_p,
\end{equation}
with the redshift factor, which is defined by the following relation
\begin{equation}
g^{s} \equiv \frac{U^{s}_\text{ob}}{U^{s}_{\text{em}}}.
\end{equation}
In the above equation ${U^s }_{\text{ob}}$ is used to define the observed photon frequency and ${U^s }_{\text{em}}$ represents the intrinsic photon
frequency \cite{Ding:2015kba}. Under the assumption that the radiation emitted is monochromatic, the rest-frame frequency $U_t$ helps us to define the emissivity as follows:
\begin{equation}\label{42}
j({U^s }_{\text{em}}) \propto \delta({U^s }_{\text{em}} -
U_t)r^{-2},
\end{equation}
where the radial emission profile is assumed to follow the dependence $r^{-2} $ 
\cite{Bambi:2013nla}. Further, the proper length measured in the emitter's
rest frame can be determined by using the following relation:
\begin{equation}\label{43}
dl_p = \sqrt{ f(r)^{-1}dr^2 + r^2 d\phi^2} = \sqrt\frac{{{1} +
\frac{r^2}{f(r)} \left( \frac{d\phi}{dr} \right)^2}}{f(r)} dr.
\end{equation}
On Substitution Eq. (\ref{43}) into Eq. (\ref{41}), one can get the
intensity as measured by a distant observer, defined as
\begin{equation}\label{44}
I(U^{s}_\text{ob}) = \int_\gamma \frac{f(r)}{r^{2}}
\sqrt\frac{{{r^2 - b^2 f(r)} + {b^2 f(r)} }}{{r^2 - b^2 f(r)}}dr.
\end{equation}

\begin{figure*}
\centering
\includegraphics[width=0.95\textwidth,keepaspectratio]{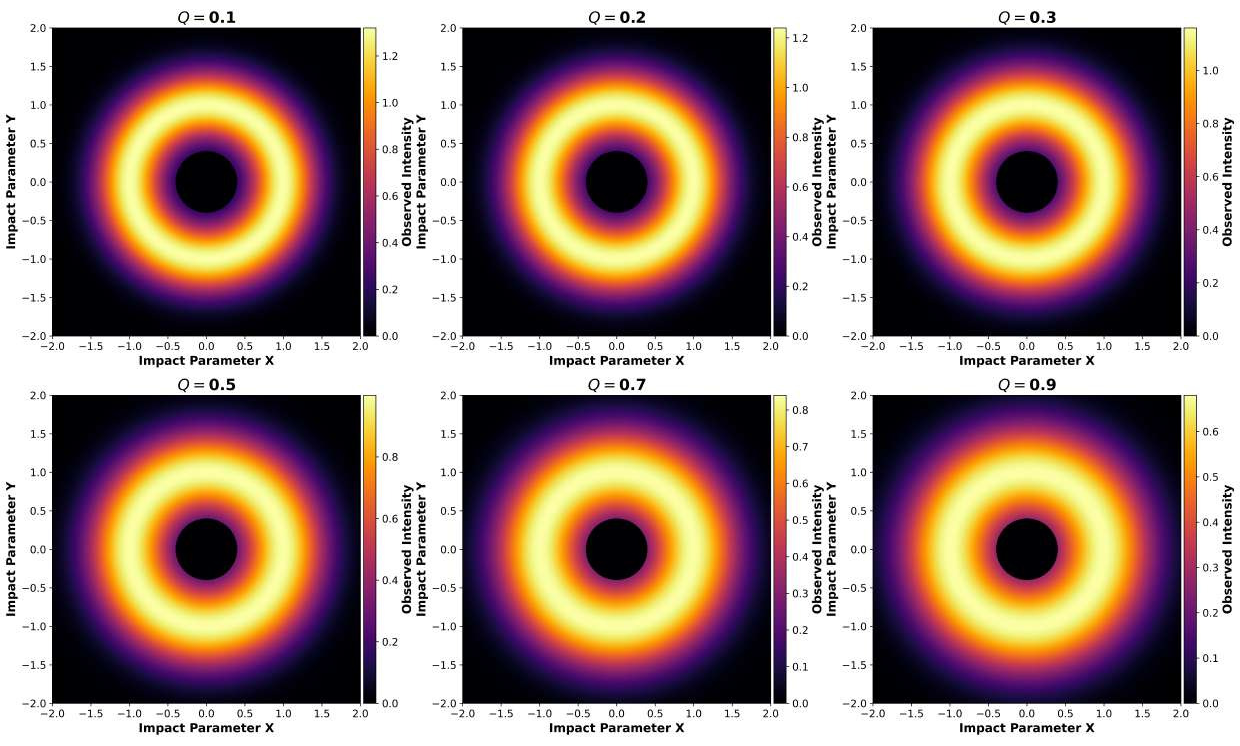}
\caption{The observed intensities around confining NED BHs with static spherical accretion for different values of $Q$.}
\label{dze1}
\end{figure*}

\begin{figure*}
\centering
\includegraphics[width=0.95\textwidth,keepaspectratio]{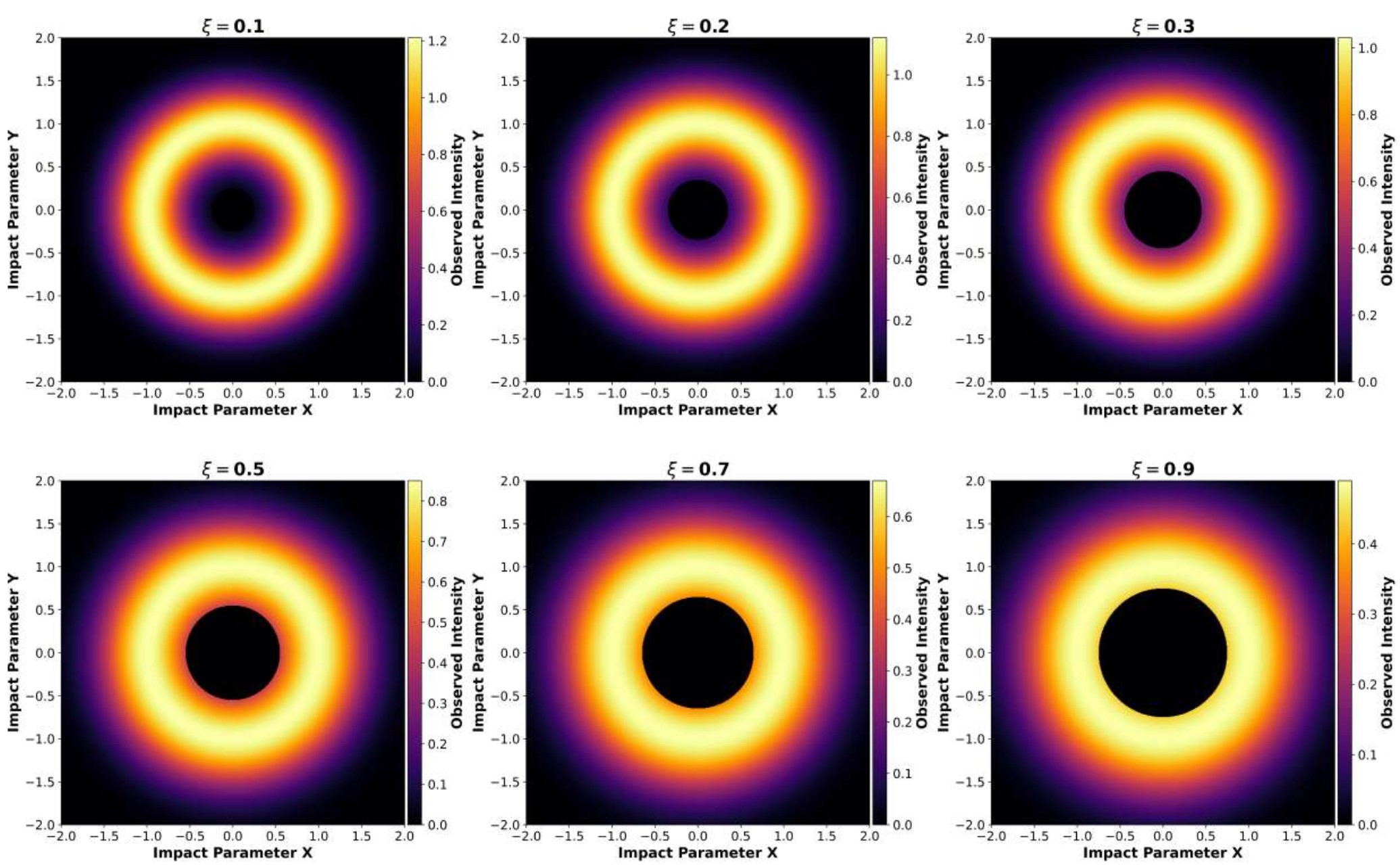}
\caption{The observed intensities around confining NED BHs with static spherical accretion for different values of $\xi$.}
\label{dze2}
\end{figure*}

\subsection{Numerical results}

Fig.~\ref{dze1} shows the total observed intensities radiated from the static spherical accretion around a confining NED BH, as a function of the impact parameter. Each plot corresponds to a different behavior against the different values of charge $Q$, for this scenario the parameter $\xi$ is fixed at $0.22$. From left to right in both rows of Fig.~\ref{dze1}, the shadow size significantly decreases, showing that the gravitational impact of the BH is strongly dependent on the parameter $Q$. At lower values of $Q$, the gravitational effects are stronger, resulting in a larger shadow. As the parameter $Q$ increases, the size of the BH shadow is reduced, indicating a weakening of the gravitational influence in the modified confining NED framework. This behavior demonstrates that the NED coupling constant 
$\xi$ plays a crucial role in determining the effective gravitational strength of the BH. The higher values of  $\xi$ diminish the gravitational attraction, resulting in smaller shadow radii, while the lower values of $\xi$ enhance the gravitational effects and produce larger shadows that closely resemble those of Schwarzschild-type BH.

Fig.~\ref{dze2} also describes the total observed intensities radiated from the static spherical accretion around a confining NED BHs, as a function of the impact parameter for the other confining NED parameter involved $\xi$. Each plot corresponds to a different value of charge $\xi$, while the parameter $Q$ is fixed at $0.1$.From left to right, the shadow radius increases noticeably, demonstrating that the gravitational influence of the BH is highly sensitive to the parameter $\xi$. For lower values of  $\xi$, the gravitational field is comparatively weaker, producing a smaller shadow. As  $\xi$ increases, the shadow becomes progressively larger, indicating an enhancement of gravitational effects within the modified confining NED framework. Consequently, the coupling constant  $\xi$effectively regulates the gravitational strength of the BH, with larger values of  $\xi$ amplifying the gravitational attraction and resulting in an expanded shadow size.

\section{Comparative Study of BHL Accretion in Confining NED and Schwarzschild Spacetimes}
\label{isec9}

In the previous sections, we established the theoretical framework for Schwarzschild-like asymptotic magnetic BHs in the context of confining NED. In doing so, we first derived the spacetime geometry of the confining NED BH and then analytically investigated its thermodynamic and orbital properties. Although these analytic calculations reveal the fundamental nature of the background spacetime and its characteristic frequencies, they do not directly describe the physical structures that emerge from the interaction between matter and the BH spacetime. To overcome this limitation, support the theoretical predictions with numerical evidence, and bridge the gap between theory and astrophysical observations, we model the mass accretion process around the confining NED magnetic BH using fully relativistic hydrodynamical simulations based on the BHL mechanism. These numerical simulations enable us to explicitly track the time-dependent behavior of the BH-matter system, directly analyze the formation of shock cones and waves, compute the evolution of the mass accretion rate, and extract the QPO frequencies generated by oscillating shock structures.

Modeling mass accretion via the BHL mechanism around BHs leads to the formation of shock cones and effective cavities in a wide class of spacetimes, including classical BHs such as Schwarzschild and Kerr \cite{Donmez:2010sx,Koyuncu:2014nga}, as well as BHs arising from modified gravity theories \cite{Donmez:2024eff,Donmez:2024luc,Donmez:2024gqb,Donmez:2025piv,Mustafa:2025mkc}. Revealing the physical properties of these structures and their dependence on model parameters provides a direct method for testing gravity through astrophysical observations. The BHL mechanism describes the theoretical framework in which matter flows supersonically toward the BH, naturally producing a shock cone structure. As matter falls toward the BH from upstream infinity, the strong gravitational field of the compact object focuses the flow, causing mass accretion near the BH and leading to the formation of a downstream shock cone. Within this narrow region, matter becomes compressed, forming a dense, elongated shock structure. The resulting shock cone behaves as a natural resonant cavity in which trapped density and pressure modes grow and interact. As density and pressure fluctuations inside the cavity are amplified and coupled through nonlinear interactions, oscillatory modes develop within the cone. These oscillations drive time-dependent accretion behavior and give rise to QPOs. Thus, the shock cone not only determines the efficiency of mass capture, but also provides a direct physical mechanism responsible for accretion variability and oscillatory phenomena in strong gravitational fields.

After matter begins to fall supersonically from the upstream region toward the BH via the BHL mechanism, the shock cone forms and reaches an approximately steady state around $t \simeq 2000M$. Beyond this time, the shock cone formed in the downstream region begins to exhibit instabilities depending on the governing parameters of the model. This time evolution and subsequent variability are discussed in detail in Fig.~\ref{acc_rate}. In order to compare the response of the oscillating shock cone in the Schwarzschild spacetime and the confining NED magnetic BH geometry, and to illustrate how the density of matter trapped inside the cone changes with system parameters, Fig.~\ref{density} presents the azimuthal variation of the rest-mass density at $r = 6.11M$ near the ISCO. The density profiles are extracted late, around $t \simeq 25000M$, well after the shock cone has fully developed and stabilized. As shown in Fig.~\ref{density}, in the Schwarzschild case the shock cone forms in the downstream region approximately within the angular range $-\pi/6 \leq \phi \leq \pi/6$, with the maximum density sharply localized near $\phi = 0$. In this case, the Schwarzschild spacetime does not introduce any additional confinement effects beyond classical gravitational focusing. In contrast, for the confining NED magnetic BH, the density of matter trapped inside the shock cone increases significantly, whereas the angular boundaries of the shock remain approximately unchanged. As the nonlinear parameter $\xi$ increases, more matter is accreted toward the BH, leading to a substantial enhancement of the density inside the cone. The resulting shock structure becomes notably more coherent and organized than in the Schwarzschild case, which in turn produces a more periodic oscillatory behavior, as discussed in Fig.~\ref{acc_rate}. However, variations in the magnetic charge parameter $Q$ do not produce any discernible modification in the shock cone geometry or its physical properties. Thus, charge-induced effects remain subdominant, whereas the NED parameter $\xi$ plays a dominant role in shaping the accretion dynamics. This demonstrates that confinement effects arising from NED are much more influential than magnetic charge in determining the structure and behavior of the shock cone.

The behavior of the shock cone shown in Fig.~\ref{density} provides a direct hydrodynamical validation of the confinement effects predicted analytically in Figs.~\ref{fig:metric_function}--\ref{lensplasma}. Theoretical calculations demonstrated that the nonlinear parameter $\xi$ modifies the thermodynamic properties, photon deflection angle, and optical geometry in the near-horizon region of the BH. The same parameter exhibits an equivalent influence when the matter becomes trapped inside the shock cone. As $\xi$ increases, the spacetime becomes more confining, leading to increased matter accretion inside the cone, stronger density contrast, and the formation of a more coherent structure near the ISCO. In contrast, no noticeable changes are observed in the density contrast or the shock cone geometry when the magnetic charge $Q$ is varied, indicating that charge-induced corrections do not play an effective role in shaping the dynamics. These results demonstrate that confinement is not merely a geometric artifact but a physically measurable phenomenon. The strong agreement between theoretical predictions and numerical results further establishes a direct connection between the NED parameter $\xi$ and the observable dynamics of accretion around the BHs.

\begin{figure*}[!ht]
  \vspace{1cm}
  \center
  \psfig{file=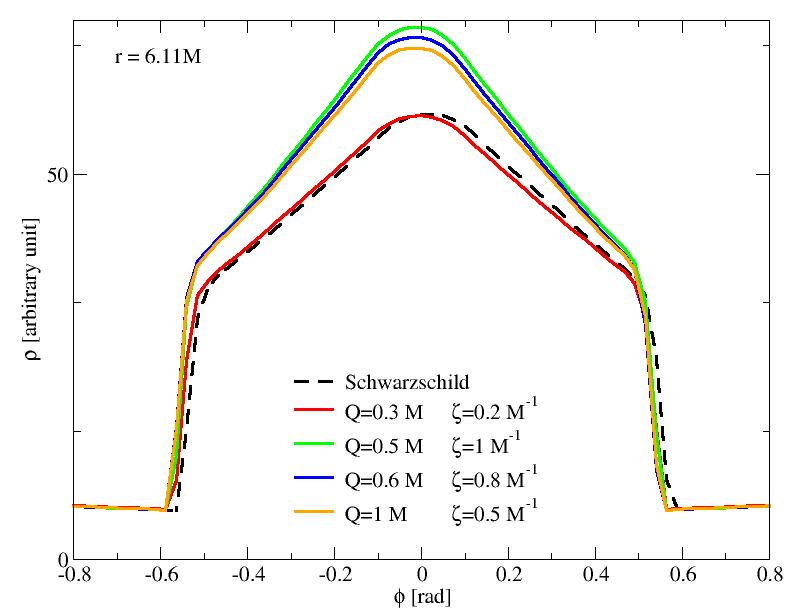,width=15.0cm,height=14.0cm}
\caption{The azimuthal variation of the rest-mass density of the shock cone formed by BHL accretion around the confining NED magnetic BH is shown at $r = 6.11M$ near the ISCO. As the nonlinear parameter $\xi$ increases, a systematic enhancement in the amount of matter trapped inside the shock cone is observed, indicating stronger confinement. Despite this increase in density, the shock locations that define the boundaries of the cone remain nearly unchanged. In contrast, the magnetic charge parameter $Q$ does not exhibit a distinguishable influence on the morphology of the shock cone. This demonstrates that nonlinear confinement effects dominate over charge-induced corrections in shaping the shock cone structure.}
\vspace{1cm}
\label{density}
\end{figure*}

Figure~\ref{acc_rate} shows the temporal evolution of the mass accretion rate measured at $r = 6.11M$ near the ISCO for different BH models. The variation of the accretion rate is presented for both the Schwarzschild and confining NED magnetic BHs, where the formation of a shock cone through the BHL mechanism plays a central role. In the Schwarzschild background, the mass accretion rate fluctuates around a relatively low mean value of approximately $\langle dM/dt \rangle \simeq 335$ (arbitrary units). The variability pattern is largely irregular, broadband, and turbulence-driven, reflecting a shock cone structure in which strong nonlinear mode couplings dominate the dynamics. In contrast, confining NED models exhibit substantially higher average accretion rates as the nonlinear parameter $\xi$ increases. For values of $\xi$ in the range $\xi = 0.5$--$1~M^{-1}$, the accretion rate varies between $\langle dM/dt \rangle \simeq 430$ and $490$, corresponding to an enhancement of nearly $40\%$ relative to the Schwarzschild case. This systematic increase directly indicates that the increase $\xi$ deepens the effective gravitational potential and significantly strengthens post-shock compression. At the same time, the temporal structure of the accretion flow evolves from stochastic behavior in the Schwarzschild geometry to a more coherent and quasi-periodic pattern in the confining NED spacetime. For higher values of $\xi$, the accretion rate displays clearly identifiable modulation cycles, as shown in Fig.~\ref{acc_rate}, indicating that nonlinear corrections promote the formation of a sharply defined hydrodynamic cavity in the downstream region.

As discussed earlier, the magnetic charge $Q$ does not affect the rest-mass density trapped inside the shock cone. However, in the case of the mass accretion rate, an additional effect emerges. Although the average accretion rate remains controlled by $\xi$ and does not show systematic dependence on $Q$, the amplitude of the oscillation decreases as $Q$ increases. This indicates that charge-induced corrections influence the temporal variability of the flow without altering the overall capture efficiency. In particular, increasing $Q$ suppresses the magnitude of time-dependent fluctuations while preserving the mean accretion rate. This behavior provides direct evidence for improved damping of nonlinear perturbations and suppression of large-scale shock distortions. Stronger confinement limits excursions in the shock geometry and preferentially selects low-order eigenmodes in the cavity. Consequently, the numerical simulations demonstrate that the accretion dynamics is primarily governed by the nonlinear confinement parameter $\xi$, while the magnetic charge $Q$ acts as a secondary stabilizing parameter in the vicinity of the BH.

In general, the numerical results establish a strong correspondence between geometric confinement and hydrodynamic behavior. Observable accretion signatures such as luminosity enhancement and QPO coherence can thus be directly linked to the underlying spacetime structure, providing a method for testing NED through astrophysical observations.

\begin{figure*}[!ht]
  \vspace{1cm}
  \center
  \psfig{file=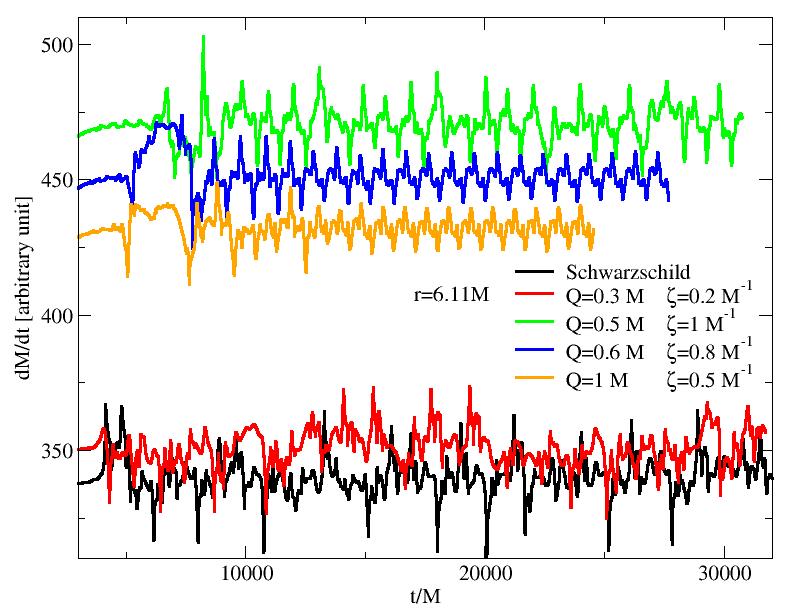,width=15.0cm,height=14.0cm}
\caption{The time evolution of the mass accretion rate evaluated at $r = 6.11M$ near the ISCO around the confining NED magnetic BH is shown after the shock cone has reached its steady state. As the nonlinear parameter $\xi$ increases, the amount of matter accreted inside the shock cone grows, which leads to a systematic increase in the accretion rate. In contrast, the magnetic charge parameter $Q$ is found to have a negligible influence on the mass accretion rate. In addition, when compared with the Schwarzschild case, the intensity and dynamical behavior of the accretion rate oscillations are significantly altered with increasing $\xi$. While the oscillation amplitude decreases, the accretion process exhibits a more periodic and coherent variability pattern for larger values of $\xi$.}
\vspace{1cm}
\label{acc_rate}
\end{figure*}

\section{Spectral Diagnostics and QPO Formation in NED Accretion Flows}
\label{isec10}

Figure~\ref{QPO_NED} presents the results of the power spectral density (PSD) analysis of the oscillation modes trapped and excited inside the shock cone formed around the confining NED magnetic BH. In addition to interpreting the numerical results obtained from these simulations, we also performed a direct comparison with both observational constraints and the QPO frequencies previously computed for Schwarzschild BH \cite{Donmez:2024gqb,Mustafa:2025gjv}. To enable a meaningful comparison, we first analyze the Schwarzschild case shown in Fig.~\ref{QPO_Schw}, where the PSD is calculated from the mass accretion rate measured near the ISCO at $r = 6.11M$. The QPO frequencies are calculated in SI units under the assumption that the BH mass is $M = M_{\odot}$. The spectrum exhibits a broadband frequency structure. Two dominant peaks appear at approximately $7.9$ Hz and $29$ Hz, which are interpreted as the result of the excitation of the fundamental oscillation modes inside the shock cone. These peaks are attributed to oscillations of the cavity formed within the shock cone in the radial direction (compression-rarefaction along the cone) and in the azimuthal direction (vortical or sloshing motion around the cone). The remaining peaks are interpreted as nonlinear coupling products arising either from self-interactions of the fundamental modes or from their mutual interactions, as well as from couplings involving already generated nonlinear modes. The broad distribution of spectral power and the absence of regular spacing or harmonic structure indicate that accretion onto the Schwarzschild BH is dominated by stronger turbulence and more chaotic dynamics.

\begin{figure*}[!ht]
  \vspace{1cm}
  \center
  \psfig{file=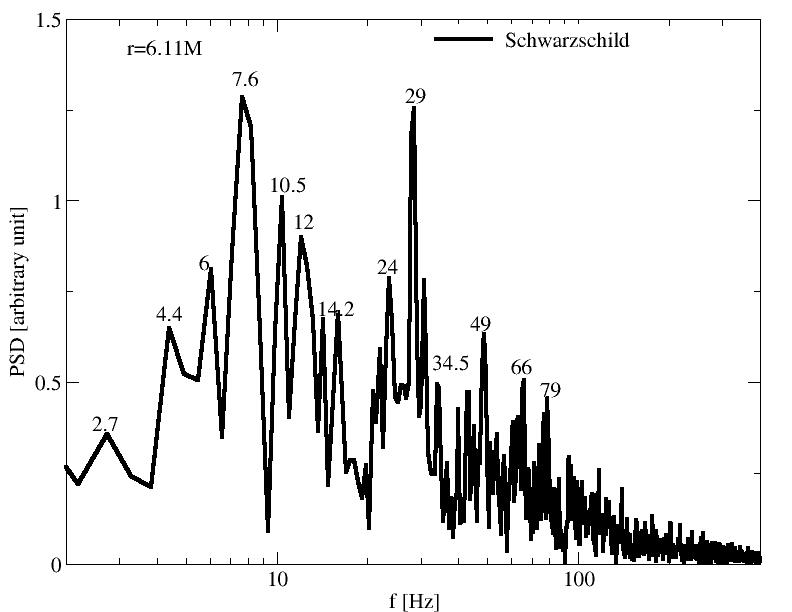,width=15.0cm,height=14.0cm}
\caption{The PSD analysis is performed using the Schwarzschild mass accretion rate shown in Fig.~\ref{acc_rate}, measured at $r = 6.11M$ near the ISCO. The fundamental modes trapped inside the oscillating shock cone and their nonlinear couplings are revealed through this analysis. Due to the complex and strongly dynamical behavior of the oscillating shock cone, both radial and azimuthal fundamental modes and their nonlinear interactions give rise to QPO frequencies in the PSD spectrum. It is observed that two dominant peaks with maximum amplitude appear at two distinct frequencies.}
\vspace{1cm}
\label{QPO_Schw}
\end{figure*}

Figure~\ref{QPO_NED} presents the PSD analysis performed at $r = 6.11M$ near the ISCO using the mass accretion rates shown in Fig.~\ref{acc_rate}, in order to reveal the modes trapped inside the shock cone formed around the confining NED magnetic BH. The variation of QPO frequencies as a function of the magnetic charge $Q$ and the nonlinear parameter $\xi$ is investigated, and the numerical results explicitly demonstrate how $Q$ and $\xi$ modify the observable frequency spectrum. All frequencies are reported in SI units under the assumption of a ten-solar-mass BH ($M = 10\,M_{\odot}$).

For the model $Q = 0.3M$ and $\xi = 0.2M^{-1}$ shown in the upper-left panel of Fig.~\ref{QPO_NED}, the PSD exhibits low-frequency QPOs (LFQPOs) and moderately high-frequency QPOs (HFQPOs) in the frequency range $4.2$--$79$ Hz. Even at this relatively weak confinement strength, a clear deviation from the Schwarzschild case is visible. From the numerically extracted frequencies, the ratio $8.5:4.2 \approx 2.02$ provides strong evidence for a $2\!:\!1$ resonance. In addition, the ratios $34.7:24.7 \approx 1.40$ and $11.4:8.5 \approx 1.34$ are close to the commensurability characteristic of $3\!:\!2$ and $4\!:\!3$, respectively. These results indicate that, in addition to the fundamental oscillation modes, nonlinear mode coupling plays an active role and leads to the formation of a set of harmonically related QPOs within the shock-cone cavity. However, compared with the Schwarzschild case, the spectrum still retains some broadband features, indicating that turbulent behavior is only partially suppressed for these relatively weak confinement parameters.

When the confinement parameters are increased to $Q = 0.5M$ and $\xi = 1M^{-1}$, a substantial modification of the spectral structure is observed, as shown in the upper-right panel of Fig.~\ref{QPO_NED}. This behavior is also evident from the significantly increased mass accretion rate displayed in Fig.~\ref{acc_rate}. In this configuration, the extracted QPO frequencies span a wide range from approximately $5$ to $132$ Hz. Compared with the cases $Q = 0.3M$ and $\xi = 0.2M^{-1}$, the frequency distribution is clearly shifted toward the HFQPO regime, while the PSD still exhibits a multi-peak structure. Several frequency pairs are remarkably close to the canonical $3\!:\!2$ ratio observed in microquasars. In particular, the ratio $91:61 \approx 1.49$ is extremely close to $3\!:\!2$, while the ratios $132:91 \approx 1.45$ and $7.3:5 \approx 1.46$ also cluster near this commensurability. In addition to the emergence of more stable frequency ratios, the amplitudes of the PSD peaks increase substantially in this case. As discussed earlier, the growth of peak amplitudes is directly correlated with the increase in the density of matter trapped inside the shock-cone cavity due to increased confinement strength $\xi$. This behavior is fully consistent with the theoretical expectations. That is, stronger confinement leads to deeper effective cavities, more efficient mode trapping, and improved energy storage in the oscillation modes. As can be clearly seen by comparing this panel with the upper-left panel of Fig.~\ref{QPO_NED}, the increase in peak amplitudes significantly improves the detectability of QPO signatures. Overall, the observed frequency shifts and increased coherence indicate that stronger confinement stabilizes the shock cone and preferentially excites a smaller number of dominant and more coherent oscillation modes within the effective cavity.

In the lower-left panel of Fig.~\ref{QPO_NED}, corresponding to the models $Q = 0.6M$ and $\xi = 0.8M^{-1}$, the PSD spectrum exhibits significantly sharper and more isolated peaks. The extracted QPO frequencies span a wide range from approximately $3.2$ to $190$ Hz, with two dominant peaks forming near $42$ and $63$ Hz. The ratio $63:42 = 1.50$ reproduces a remarkably clear frequency ratio $3\!:\!2$, in excellent agreement with HFQPO observations reported in the BH X-ray binaries. In addition, HFQPO pairs appear in $147:106 \approx 1.39$ and $190:106 \approx 1.79$, revealing the presence of extra commensurabilities and nonlinear couplings within the oscillation spectrum. The emergence of multiple high-frequency peaks indicates a ladder of nonlinear couplings and higher-order oscillation modes inside the effective cavity. The appearance of harmonically related peaks in the high-frequency regime strongly supports the interpretation that nonlinear electromagnetic confinement introduces a frequency selection mechanism analogous to resonance models in relativistic disk dynamics.

In the lower-right panel of Fig.~\ref{QPO_NED}, corresponding to the cases $Q = 1M$ and $\xi = 0.5M^{-1}$, both LFQPOs and HFQPOs are observed within the wide frequency range of approximately $2.8$--$156$ Hz. The shift of the dominant oscillations toward the high-frequency band ($\sim 50$--$200$ Hz) demonstrates that the increasing magnetic charge directly elevates the characteristic QPO frequencies. For this configuration, several frequency ratios emerge that may be directly compared with observational data. The ratio $67.4:47 \approx 1.43$ suggests a resonance near $3\!:\!2$, while $47:21.8 \approx 2.16$ and $90:47 \approx 1.91$ appear close to the $2\!:\!1$ ratio. These commensurabilities indicate enhanced nonlinear coupling between oscillation modes inside the effective cavity.

In contrast to the behavior of HFQPOs, LFQPO amplitudes are found to be more strongly influenced by the nonlinear confinement parameter $\xi$. For lower values of $\xi$, the LFQPOs appear with significantly larger amplitudes and are therefore more easily detectable. For example, in the upper-left panel of Fig.~\ref{QPO_NED} corresponding to $\xi = 0.2M^{-1}$, LFQPOs are clearly dominant with strong amplitudes. However, as $\xi$ increases, the LFQPO amplitudes gradually decrease, and for $\xi = 0.8M^{-1}$ and $1M^{-1}$ their detectability is substantially reduced.

However, increasing the magnetic charge $Q$ shifts the spectral power toward the HFQPO regime. This indicates that even in the absence of BH spin, magnetic charge in combination with NED provides a natural mechanism for generating HFQPOs. Consequently, magnetic charge and nonlinear electrodynamic corrections emerge as an alternative frequency-setting mechanism that can reorganize the QPO spectrum independently of frame-dragging effects.

When the behavior of QPO frequencies obtained around the confining NED magnetic BH (Fig.~\ref{QPO_NED}) is compared with the Schwarzschild model (Fig.~\ref{QPO_Schw}), the differences become particularly striking. In the Schwarzschild case, the PSD analysis reveals a large number of LFQPOs and moderately strong HFQPOs, with frequency ratios close to $2\!:\!1$ and $3\!:\!2$. However, these ratios are embedded in a broadband turbulent background. This indicates that in Schwarzschild spacetime the shock cone excites many interacting modes, most of which possess comparable amplitudes, thereby producing a chaotic peak structure. In contrast, in the confining NED BH geometry, broadband turbulence is suppressed, and the system evolves toward more periodic oscillatory behavior. The PSD analysis demonstrates the emergence of fewer but significantly stronger peaks, accompanied by a shift of characteristic frequencies toward the HFQPO regime. These features indicate that the shock cone operates as a highly efficient resonant cavity whose eigenfrequencies are regulated by the NED parameter $\xi$ and the magnetic charge $Q$. The confinement parameter $\xi$ primarily controls the amplitude and coherence of the oscillatory modes, while the magnetic charge $Q$ governs the elevation of frequencies and determines the overall spectral range. This clear separation of roles highlights how nonlinear electrodynamic confinement reshapes both the temporal structure of the accretion flow and the observable QPO spectrum.

Compared with observations, the results obtained here for the non-rotating confining NED BH are particularly significant. This is because stable $3\!:\!2$ HFQPO pairs are commonly reported in BH binary systems such as GRS~1915+105 \cite{Belloni:2019sot,Zhang:2022epl}, GRO~J1655-40 \cite{Stuchlik:2016brp,Chakrabarti:2005bd}, and XTE~J1550-564 \cite{Wang:2008xda}. These frequency ratios are generally interpreted within the framework of epicyclic resonance models in Kerr spacetime. However, the PSD analysis presented for the confining NED BH reveals an alternative mechanism capable of reproducing the same observational features. In this framework, electromagnetic confinement and magnetic charge provide a natural explanation for the emergence of stable HFQPO pairs even in the absence of frame-dragging effects. Consequently, the confining NED magnetic BH model offers a viable alternative to rotating BH scenarios for interpreting QPO observations.

\begin{figure*}[!ht]
  \vspace{1cm}
  \center
  \psfig{file=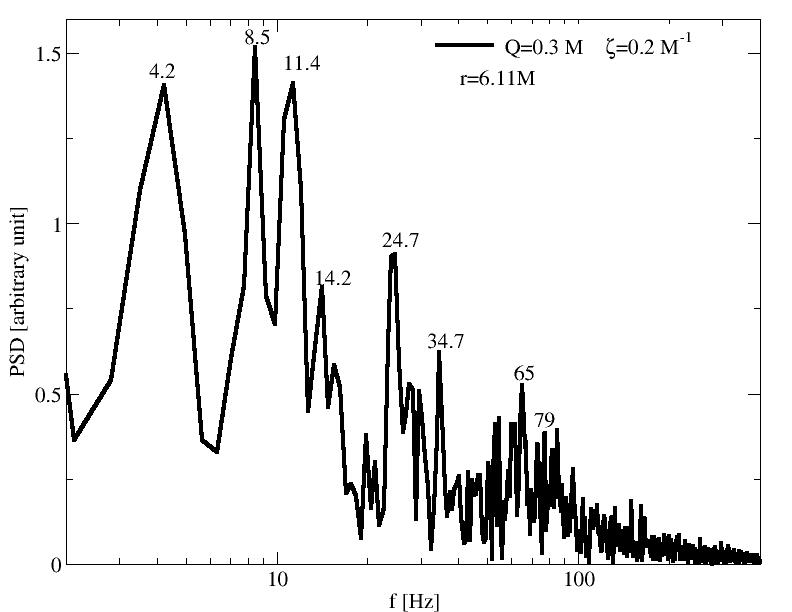,width=7.5cm,height=7.0cm}
  \psfig{file=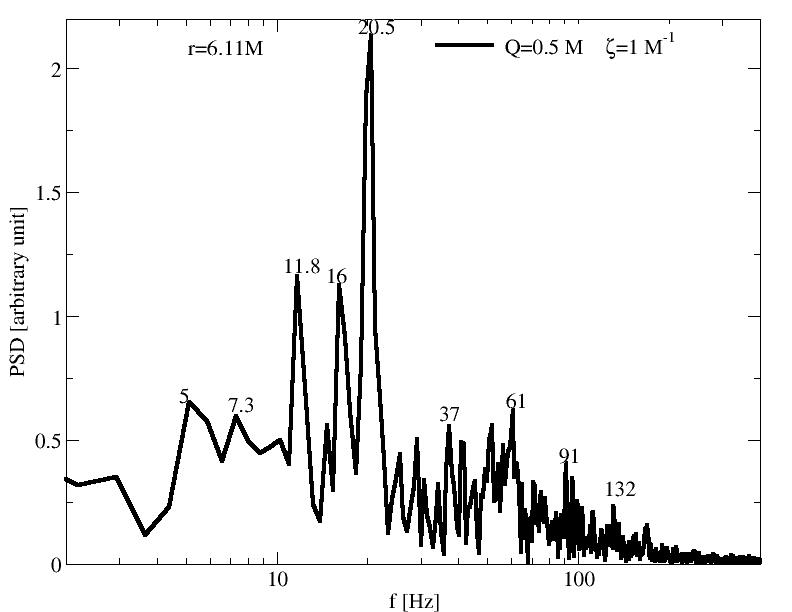,width=7.5cm,height=7.0cm} \\
  \psfig{file=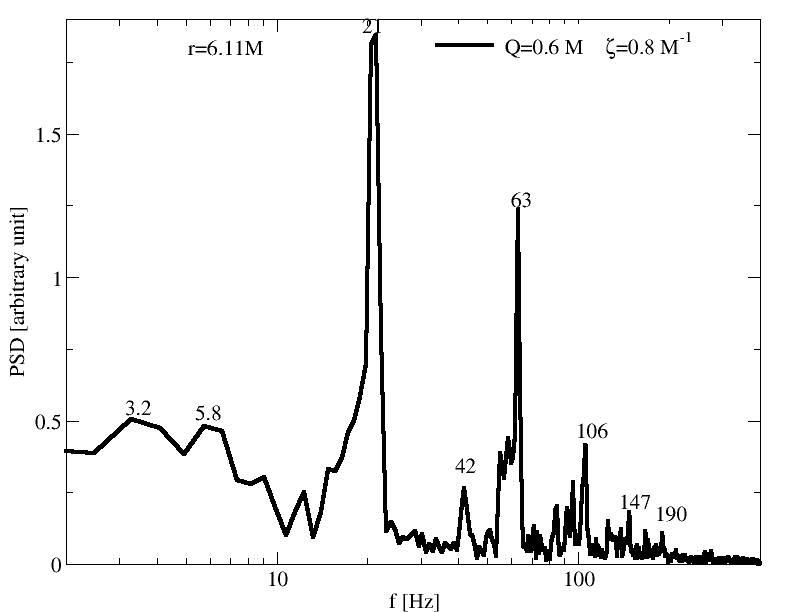,width=7.5cm,height=7.0cm}
  \psfig{file=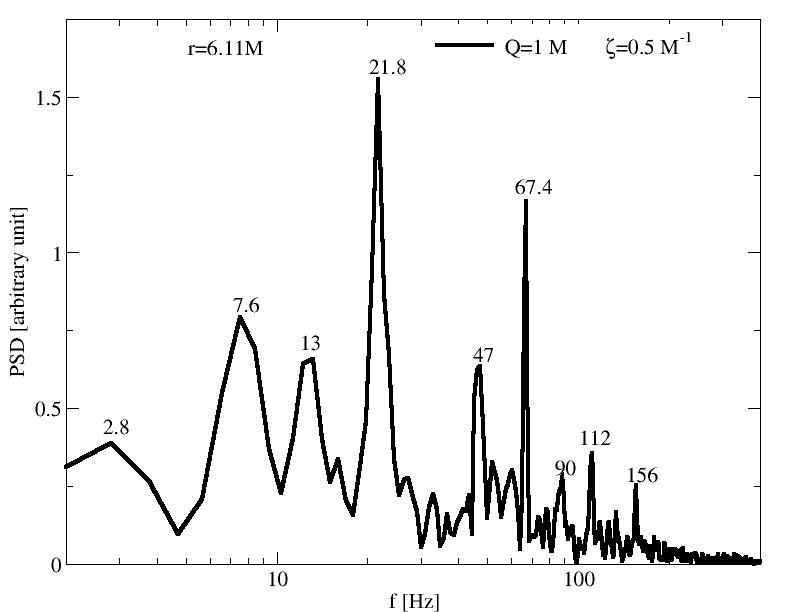,width=7.5cm,height=7.0cm} \\  
\caption{As in Fig.~\ref{QPO_Schw}, a PSD analysis is also performed here using the mass accretion rates shown in Fig.~\ref{acc_rate}. However, in this case the analysis is carried out for the oscillating shock cone formed around the confining NED magnetic BH. Each panel displays the QPO frequencies obtained for different values of the magnetic charge $Q$ and nonlinear parameter $\xi$. It is observed that, as $\xi$ increases, the QPO frequencies associated with the trapping of fundamental modes inside the effective cavity are significantly modified, accompanied by an enhancement in the maximum amplitude of the PSD peaks. In addition, the system exhibits an increased number of HFQPOs in the power spectrum.}
\vspace{1cm}
\label{QPO_NED}
\end{figure*}

\section{Conclusion} \label{isec11}

In this work, we carried out a detailed investigation of the confining NED BH, a static, spherically symmetric spacetime arising from Einstein gravity minimally coupled to a nonlinear electromagnetic field that reproduces Maxwell theory in the strong-field regime while introducing confinement-like corrections at large distances. This inverted behavior—opposite to classical NED models such as Born-Infeld or Euler-Heisenberg—produces a geometry that is asymptotically Schwarzschild but carries a characteristic $Q^3/(9\xi^2 r^4)$ correction to the metric function, as shown in Eq.~\eqref{metricFun}. The absence of the standard RN term $Q^2/r^2$ reflects the confining nature of the electromagnetic sector and leads to observable signatures that distinguish this model from both Schwarzschild and RN BHs.

We began in Sec.~\ref{isec2} by deriving the spacetime geometry and analyzing the horizon structure. The metric function $f(r)$ depends on the Schwarzschild mass $M$, the magnetic charge $Q$, and the NED parameter $\xi$, with the ADM mass receiving a logarithmic correction from the electromagnetic self-energy according to Eq.~(7). Table~\ref{tab:NED_horizons} presented a survey of horizon configurations across the parameter space $(Q, \xi)$, revealing transitions between NE BH solutions with two distinct horizons, Ext configurations with degenerate horizons, and NS spacetimes where no horizon exists. Figure~\ref{fig:metric_function} illustrated these different regimes through the radial profile of $f(r)$, showing how increasing $Q$ at fixed $\xi$ pushes the system toward extremality, while increasing $\xi$ restores the structure of the horizon by enhancing the mass of ADM. The 3D embedding diagrams in Fig.~\ref{fig:embedding_panels} visualized the spatial geometry, demonstrating that the EH radius grows monotonically with both $Q$ and $\xi$, and that the gravitational funnel becomes wider and shallower for larger horizons.

The weak-field gravitational lensing analysis in Sec.~\ref{isec3} employed the GBT to compute the deflection angle for photons passing near the confining NED BH. The vacuum result, given in Eq.~(13), showed that the NED corrections introduce additional inverse power-law terms scaling as $Q^3/(\xi b^4)$ and $Q^6/(\xi^2 b^8)$, all carrying negative signs that reduce the total bending compared to a Schwarzschild BH of equal ADM mass. Figure~\ref{lens} showed the deflection angle as a function of the impact parameter and $\xi$, confirming the expected $1/b$ falloff at large distances. When plasma effects were incorporated through a frequency-dependent refractive index, deflection acquired chromatic corrections proportional to the plasma parameter $\delta = \omega_e^2/\omega_0^2$, as shown in Eq.~(21) and Fig.~\ref{lensplasma}. These results established that multi-frequency lensing observations could in principle disentangle gravitational and dispersive contributions, providing constraints on the NED parameters.

Section~\ref{isec4} extended lensing analysis to include axion-photon coupling in a magnetized plasma environment. The refractive index acquired corrections from plasma dispersion and axion dynamics, with a resonance structure appearing when the photon frequency approaches the axion mass $\omega_\varphi$. The resulting deflection angle, given in Eq.~(30), exhibited a distinctive energy dependence that differs from pure plasma chromatic effects and could serve as an observational signature of axionic DM in the vicinity of magnetized compact objects.

The gravitational redshift analysis in Sec.~\ref{isec5} demonstrated that photons escaping from the confining NED BH experience a frequency shift determined by the metric function at the emission point, as expressed in Eq.~\eqref{eq:z_infty_exact_NED}. The weak-field expansion in Eq.~\eqref{eq:z_linear_NED} separated the redshift into a Schwarzschild-type term $M_{\text{ADM}}/r_e$ and an NED correction $-Q^3/(18\xi^2 r_e^4)$ that reduces the total shift due to its negative sign. Figure~\ref{red} showed that the redshift increases steeply near the EH and varies with $\xi$ through the competition between the enhanced ADM mass and the suppressed NED correction term. The $r^{-4}$ scaling of the NED contribution offers a potential method to distinguish this geometry from standard GR predictions using X-ray spectroscopy of emission lines from matter orbiting close to the horizon.

The thermodynamic properties examined in Sec.~\ref{isec6} revealed that the confining NED BH exhibits nontrivial thermal behavior in the extended phase space. The JTE coefficient $\mu_J$, given in Eq.~(41) changes sign at an inversion point that depends on the NED parameters, separating the cooling ($\mu_J > 0$) and heating ($\mu_J < 0$) regimes during isenthalpic expansion. Figure~\ref{jte} displayed the variation of $\mu_J$ with horizon radius and $\xi$, showing that larger values of the NED parameter shift the inversion curve to smaller radii. The heat capacity $C_P$ in Eq.~(43) exhibited a divergence at a critical radius marking a second-order phase transition between thermally stable and unstable configurations, as illustrated in Fig.~\ref{cc}. These thermodynamic signatures complement the lensing and redshift results, providing additional probes of the confining electromagnetic structure.

The stability analysis of circular geodesics in Sect.~\ref{isec7} showed that the radius of the photon sphere satisfies a quartic equation, Eq.~\eqref{eq:photon_sphere_equation}, rather than the cubic or linear relations found in the RN and Schwarzschild spacetimes, respectively. The Lyapunov exponent characterizing the instability of null circular orbits was derived in closed form in Eq.~(53), with explicit dependence on both $Q$ and $\xi$ through the combination $Q^3/\xi^2$. This exponent controls the decay rate of BH ringdown signals in the eikonal approximation and the width of the photon ring structure, offering potential observational tests through QNM spectroscopy and high-resolution shadow imaging.

Section~\ref{isec8} computed the BH shadow for the confining NED geometry. The shadow radius, given by the critical impact parameter $b_{\rm ph} = r_{\rm ph}/\sqrt{f(r_{\rm ph})}$ in Eq.~(59), increases with $\xi$ due to the enhanced ADM mass. Figure~\ref{fig:shadow_fullpage} presented synthetic shadow images for three values of the coupling parameter, demonstrating that a larger $\xi$ produces broader shadows while maintaining circular symmetry. The connection between the shadow boundary and the Lyapunov exponent was emphasized: the width of the photon ring encircling the shadow encodes information about orbital instability that could be extracted from next-generation EHT observations.

The numerical simulations of BHL accretion in Sec.~\ref{isec9} provided direct hydrodynamical validation of the confinement effects predicted analytically. Figure~\ref{density} showed that the density of the rest of the mass inside the shock cone increases significantly with the NED parameter $\xi$, while the angular boundaries of the cone remain nearly unchanged. The mass accretion rate, displayed in Fig.~\ref{acc_rate}, exhibited a $\sim 40\%$ enhancement relative to the Schwarzschild case for $\xi$ in the range $0.5$--$1~M^{-1}$. While $\xi$ controlled the mean accretion rate and density, the magnetic charge $Q$ influenced the temporal variability by suppressing large-scale oscillations and damping nonlinear instabilities. These results demonstrated that confinement is not merely a geometric artifact but a physically measurable phenomenon with direct consequences for accretion dynamics.

The PSD analysis in Sec.~\ref{isec10} revealed that the confining NED geometry reorganizes the QPO frequency spectrum compared to Schwarzschild spacetime. In the Schwarzschild case (Fig.~\ref{QPO_Schw}), broadband turbulence dominated the spectral structure with weakly isolated peaks. In contrast, the confining NED BH (Fig.~\ref{QPO_NED}) produced fewer but significantly stronger peaks, with characteristic frequencies shifting toward the HFQPO regime ($50$--$200$ Hz). Stable $3\!:\!2$ and $2\!:\!1$ frequency ratios emerged consistently across different parameter configurations, matching observational results from BH X-ray binaries such as GRS~1915+105 \cite{Belloni:2019sot,Zhang:2022epl}, GRO~J1655-40 \cite{Stuchlik:2016brp,Chakrabarti:2005bd}, and XTE~J1550-564 \cite{Wang:2008xda}. The parameter $\xi$ mainly governed the amplitude and coherence of the oscillation modes, while $Q$ controlled the frequency elevation. This separation of roles established that electromagnetic confinement and magnetic charge together provide a spin-independent mechanism for generating HFQPOs, offering a viable alternative to Kerr-based epicyclic resonance models.

Looking forward, several directions merit further investigation. First, extending the analysis to rotating (Kerr-like) confining NED BHs would allow direct comparison with spin-based QPO models and enable constraints from combined mass-spin measurements. Second, computing the full QNM spectrum beyond the eikonal approximation would test whether the $\xi$-dependent Lyapunov exponent correctly predicts the ringdown waveforms detectable by gravitational wave observatories. Third, incorporating realistic accretion disk models with radiative transfer would produce synthetic spectra and images that can be compared directly with EHT observations of M87* and Sgr~A*. Fourth, the axion-plasma lensing formalism developed here could be applied to magnetar environments where strong magnetic fields and potential axion backgrounds coexist, providing new probes of beyond-Standard Model physics. Finally, the thermodynamic phase structure suggests the possibility of BH phase transitions that could leave imprints on the time-domain behavior of accreting systems, a connection that deserves a dedicated study. These extensions would further establish the confining NED BH as a testable theoretical framework with direct relevance to current and upcoming astrophysical observations.

}

%%%%%%%%%%%%%%%%%%%%%%%%%%%%%%%%%%%%%%%%%%%%%%%%%%%%%%%%%%%%%%%%%%%%%%%%

%%%%%%%%%%%%%%%%%%%%%%%%%%%%%%%%%%%%%%%%%%%%%%%%%%%%%%%%%%%%%%%%%%

\newpage
\acknowledgments 
All numerical simulations were performed using the Phoenix High
Performance Computing facility at the American University of the Middle East (AUM), Kuwait. \.{I}.~S. and E. S. gratefully acknowledge EMU, T\"{U}B\.{I}TAK, ANKOS, and SCOAP3 for supporting networking activities within COST Actions CA21106, CA21136, CA22113, CA23115, and CA23130.

\bibliography{ref}

@article{mazharimousavi2024confinement,
  title={Confinement and nonlinear electrodynamics: Asymptotic Schwarzschild charged black hole},
  author={Mazharimousavi, S Habib},
  journal={Physics of the Dark Universe},
  volume={43},
  pages={101413},
  year={2024},
  publisher={Elsevier}
}

@article{sucu2025astrophysical,
  title={Astrophysical reality of black hole thermodynamics and dynamics: Transformative influence of Hernquist dark matter distributions},
  author={Sucu, Erdem and Sakall{\i}, {\.I}zzet},
  journal={Physics of the Dark Universe},
  pages={102051},
  year={2025},
  publisher={Elsevier}
}

@article{Bambi:2013nla,
    author = "Bambi, Cosimo",
    title = "{Can the supermassive objects at the centers of galaxies be traversable wormholes? The first test of strong gravity for mm/sub-mm very long baseline interferometry facilities}",
    eprint = "1304.5691",
    archivePrefix = "arXiv",
    primaryClass = "gr-qc",
    doi = "10.1103/PhysRevD.87.107501",
    journal = "Phys. Rev. D",
    volume = "87",
    pages = "107501",
    year = "2013"
}

@article{Ding:2015kba,
    author = "Ding, Chikun and Wang, Anzhong and Wang, Xinwen",
    title = "{Charged Einstein-aether black holes and Smarr formula}",
    eprint = "1507.06618",
    archivePrefix = "arXiv",
    primaryClass = "gr-qc",
    doi = "10.1103/PhysRevD.92.084055",
    journal = "Phys. Rev. D",
    volume = "92",
    number = "8",
    pages = "084055",
    year = "2015"
}

@article{sucu2025charged,
  title={Charged regular black holes in quantum gravity: from thermodynamic stability to observational phenomena},
  author={Sucu, Erdem and Sakall{\i}, {\.I}zzet},
  journal={The European Physical Journal C},
  volume={85},
  number={9},
  pages={989},
  year={2025},
  publisher={Springer}
}

@article{Berti:2015itd,
    author = "Berti, Emanuele and others",
    title = "{Testing General Relativity with Present and Future Astrophysical Observations}",
    eprint = "1501.07274",
    archivePrefix = "arXiv",
    primaryClass = "gr-qc",
    doi = "10.1088/0264-9381/32/24/243001",
    journal = "Class. Quant. Grav.",
    volume = "32",
    pages = "243001",
    year = "2015"
}

@article{Johannsen:2016uoh,
    author = "Johannsen, Tim",
    title = "{Testing the No-Hair Theorem with Observations of Black Holes in the Electromagnetic Spectrum}",
    eprint = "1602.07694",
    archivePrefix = "arXiv",
    primaryClass = "astro-ph.HE",
    doi = "10.1088/0264-9381/33/12/124001",
    journal = "Class. Quant. Grav.",
    volume = "33",
    number = "12",
    pages = "124001",
    year = "2016"
}

@article{abbott2016observation,
  title={Observation of gravitational waves from a binary black hole merger},
  author={Abbott, Benjamin P and Abbott, Richard and Abbott, Thomas D and Abernathy, Matthew R and Acernese, Fausto and Ackley, Kendall and Adams, Carl and Adams, Thomas and Addesso, Paolo and Adhikari, Rana X and others},
  journal={Physical review letters},
  volume={116},
  number={6},
  pages={061102},
  year={2016},
  publisher={APS}
}

@article{abbott2016tests,
  title={Tests of general relativity with GW150914},
  author={Abbott, Benjamin P and Abbott, R and Abbott, TD and Abernathy, MR and Acernese, Fausto and Ackley, K and Adams, C and Adams, T and Addesso, Paolo and Adhikari, RX and others},
  journal={Physical review letters},
  volume={116},
  number={22},
  pages={221101},
  year={2016},
  publisher={APS}
}

@article{event2019first,
  title={First M87 event horizon telescope results. I. The shadow of the supermassive black hole},
  author={Event Horizon Telescope Collaboration and others},
  journal={arXiv preprint arXiv:1906.11238},
  year={2019}
}

@article{akiyama2022first,
  title={First Sagittarius A* Event Horizon Telescope results. I. The shadow of the supermassive black hole in the center of the Milky Way},
  author={Akiyama, Kazunori and Alberdi, Antxon and Alef, Walter and Algaba, Juan Carlos and Anantua, Richard and Asada, Keiichi and Azulay, Rebecca and Bach, Uwe and Baczko, Anne-Kathrin and Ball, David and others},
  journal={The Astrophysical Journal Letters},
  volume={930},
  number={2},
  pages={L12},
  year={2022},
  publisher={IOP Publishing}
}

@article{bonanno2000renormalization,
  title={Renormalization group improved black hole spacetimes},
  author={Bonanno, Alfio and Reuter, Martin},
  journal={Physical Review D},
  volume={62},
  number={4},
  pages={043008},
  year={2000},
  publisher={APS}
}

@article{ayon1998regular,
  title={Regular black hole in general relativity coupled to nonlinear electrodynamics},
  author={Ayon-Beato, Eloy and Garcia, Alberto},
  journal={Physical review letters},
  volume={80},
  number={23},
  pages={5056},
  year={1998},
  publisher={APS}
}

@article{heisenberg2006consequences,
  title={Consequences of dirac theory of the positron},
  author={Heisenberg, W and Euler, H},
  journal={arXiv preprint physics/0605038},
  year={2006}
}

@article{lai2001matter,
  title={Matter in strong magnetic fields},
  author={Lai, Dong},
  journal={Reviews of Modern Physics},
  volume={73},
  number={3},
  pages={629},
  year={2001},
  publisher={APS}
}

@article{hendi2014thermodynamic,
  title={Thermodynamic properties of asymptotically Reissner--Nordstr{\"o}m black holes},
  author={Hendi, Seyed Hossein},
  journal={Annals of Physics},
  volume={346},
  pages={42--50},
  year={2014},
  publisher={Elsevier}
}

@article{sucu2025dynamics,
  title={Dynamics of particles surrounding a stationary, spherically-symmetric black hole with nonlinear electrodynamics},
  author={Sucu, Erdem and Sakall{\i}, Izzet},
  journal={Physics of the Dark Universe},
  volume={47},
  pages={101771},
  year={2025},
  publisher={Elsevier}
}

@article{gursel2025thermodynamics,
  title={Thermodynamics of Einstein-Euler-Heisenberg Black Holes with Thermal Fluctuations and Nonlinear Electromagnetic Fields},
  author={Gursel, Huriye and Mangut, Mert and Sucu, Erdem},
  journal={Classical and Quantum Gravity},
  year={2025}
}

@article{bekenstein1973black,
  title={Black holes and entropy},
  author={Bekenstein, Jacob D},
  journal={Physical Review D},
  volume={7},
  number={8},
  pages={2333},
  year={1973},
  publisher={APS}
}

@article{sahan2025quantum,
  title={Quantum phase transitions of Dirac particles in a magnetized rotating curved background: Interplay of geometry, magnetization, and thermodynamics},
  author={Sahan, Nusret and Sucu, Erdem and Sucu, Yusuf},
  journal={Physics of the Dark Universe},
  pages={102005},
  year={2025},
  publisher={Elsevier}
}

@article{reynolds2003fluorescent,
  title={Fluorescent iron lines as a probe of astrophysical black hole systems},
  author={Reynolds, Christopher S and Nowak, Michael A},
  journal={Physics Reports},
  volume={377},
  number={6},
  pages={389--466},
  year={2003},
  publisher={Elsevier}
}

@article{yasir2023thermal,
  title={Thermal geometries and the Joule--Thomson expansion of modified charged and slowly rotating black holes},
  author={Yasir, Muhammad and Lining, Tong and Tiecheng, Xia and Ditta, Allah},
  journal={Frontiers in Physics},
  volume={11},
  pages={1170683},
  year={2023},
  publisher={Frontiers Media SA}
}

@article{silva2021joule,
  title={Joule--Thomson expansion in charged AdS black hole surrounded by a cosmological fluid in Rainbow Gravity},
  author={Silva, GV and Bezerra, VB and Gra{\c{c}}a, JP Morais and Lobo, IP},
  journal={Modern Physics Letters A},
  volume={36},
  number={40},
  pages={2150278},
  year={2021},
  publisher={World Scientific}
}

@article{dolan2011cosmological,
  title={The cosmological constant and black-hole thermodynamic potentials},
  author={Dolan, Brian P},
  journal={Classical and Quantum Gravity},
  volume={28},
  number={12},
  pages={125020},
  year={2011},
  publisher={IOP Publishing}
}

@article{kruglov2023magnetic,
  title={Magnetic black holes in AdS space with nonlinear electrodynamics, extended phase space thermodynamics and Joule--Thomson expansion},
  author={Kruglov, SI},
  journal={International Journal of Geometric Methods in Modern Physics},
  volume={20},
  number={01},
  pages={2350008},
  year={2023},
  publisher={World Scientific}
}

@article{kruglov2022ned,
  title={NED-AdS black holes, extended phase space thermodynamics and Joule--Thomson expansion},
  author={Kruglov, SI},
  journal={Nuclear Physics B},
  volume={984},
  pages={115949},
  year={2022},
  publisher={Elsevier}
}

@book{wald2024general,
  title={General relativity},
  author={Wald, Robert M},
  year={2024},
  publisher={University of Chicago press}
}

@article{fabian1989x,
  title={X-ray fluorescence from the inner disc in Cygnus X-1},
  author={Fabian, AC and Rees, MJ and Stella, L and White, Ne E},
  journal={Monthly Notices of the Royal Astronomical Society},
  volume={238},
  number={3},
  pages={729--736},
  year={1989},
  publisher={Oxford University Press Oxford, UK}
}

@article{pound1959gravitational,
  title={Gravitational red-shift in nuclear resonance},
  author={Pound, Robert V and Rebka Jr, Glen A},
  journal={Physical Review Letters},
  volume={3},
  number={9},
  pages={439},
  year={1959},
  publisher={APS}
}

@article{pound1960apparent,
  title={Apparent weight of photons},
  author={Pound, Robert V and Rebka Jr, Glen A},
  journal={Physical review letters},
  volume={4},
  number={7},
  pages={337},
  year={1960},
  publisher={APS}
}

@article{henneaux1984cosmological,
  title={The cosmological constant as a canonical variable},
  author={Henneaux, Marc and Teitelboim, Claudio},
  journal={Physics Letters B},
  volume={143},
  number={4-6},
  pages={415--420},
  year={1984},
  publisher={Elsevier}
}

@article{aydiner2025regular,
  title={Regular magnetically charged black holes from nonlinear electrodynamics: Thermodynamics, light deflection, and orbital dynamics},
  author={Aydiner, Ekrem and Sucu, Erdem and Sakall{\i}, {\.I}zzet},
  journal={Physics of the Dark Universe},
  pages={102164},
  year={2025},
  publisher={Elsevier}
}

@article{york1986black,
  title={Black-hole thermodynamics and the Euclidean Einstein action},
  author={York Jr, James W},
  journal={Physical Review D},
  volume={33},
  number={8},
  pages={2092},
  year={1986},
  publisher={APS}
}

@article{synge1966escape,
  title={The escape of photons from gravitationally intense stars},
  author={Synge, JL},
  journal={Monthly Notices of the Royal Astronomical Society},
  volume={131},
  number={3},
  pages={463--466},
  year={1966},
  publisher={Oxford University Press Oxford, UK}
}

@article{hioki2009measurement,
  title={Measurement of the Kerr spin parameter by observation of a compact object’s shadow},
  author={Hioki, Kenta and Maeda, Kei-ichi},
  journal={Physical Review D—Particles, Fields, Gravitation, and Cosmology},
  volume={80},
  number={2},
  pages={024042},
  year={2009},
  publisher={APS}
}

@article{cardoso2009geodesic,
  title={Geodesic stability, Lyapunov exponents, and quasinormal modes},
  author={Cardoso, Vitor and Miranda, Alex S and Berti, Emanuele and Witek, Helvi and Zanchin, Vilson T},
  journal={Physical Review D—Particles, Fields, Gravitation, and Cosmology},
  volume={79},
  number={6},
  pages={064016},
  year={2009},
  publisher={APS}
}

@article{bondi1944mechanism,
  title={On the mechanism of accretion by stars},
  author={Bondi, Hermann and Hoyle, Fred},
  journal={Monthly Notices of the Royal Astronomical Society},
  volume={104},
  number={5},
  pages={273--282},
  year={1944},
  publisher={Oxford Academic}
}

@inproceedings{hoyle1939effect,
  title={The effect of interstellar matter on climatic variation},
  author={Hoyle, Fred and Lyttleton, Raymond A},
  booktitle={Mathematical proceedings of the Cambridge philosophical society},
  volume={35},
  number={3},
  pages={405--415},
  year={1939},
  organization={Cambridge University Press}
}

@article{ruffert1994three,
  title={Three-dimensional hydrodynamic Bondi-Hoyle accretion. III. Mach 0.6, 1.4 and 10; gamma= 5/3.},
  author={Ruffert, M},
  journal={Astronomy and Astrophysics Suppl., Vol. 106, p. 505-522 (1994)},
  volume={106},
  pages={505--522},
  year={1994}
}

@article{donmez2012relativistic,
  title={Relativistic simulation of flip-flop instabilities of Bondi--Hoyle accretion and quasi-periodic oscillations},
  author={D{\"o}nmez, Orhan},
  journal={Monthly Notices of the Royal Astronomical Society},
  volume={426},
  number={2},
  pages={1533--1545},
  year={2012},
  publisher={Blackwell Science Ltd Oxford, UK}
}

@article{donmez2024bondi,
  title={Bondi-Hoyle-Lyttleton accretion around the rotating hairy Horndeski black hole},
  author={D{\"o}nmez, O},
  journal={Journal of Cosmology and Astroparticle Physics},
  volume={2024},
  number={09},
  pages={006},
  year={2024},
  publisher={IOP Publishing}
}

@incollection{van1989fourier,
  title={Fourier techniques in X-ray timing},
  author={Van der Klis, M},
  booktitle={Timing neutron stars},
  pages={27--69},
  year={1989},
  publisher={Springer}
}

@article{balart2014regular,
  title={Regular black holes with a nonlinear electrodynamics source},
  author={Balart, Leonardo and Vagenas, Elias C},
  journal={Physical Review D},
  volume={90},
  number={12},
  pages={124045},
  year={2014},
  publisher={APS}
}

@article{vegetti2024strong,
  title={Strong gravitational lensing as a probe of dark matter},
  author={Vegetti, S and Birrer, S and Despali, G and Fassnacht, CD and Gilman, D and Hezaveh, Y and Perreault Levasseur, L and McKean, JP and Powell, Devon M and O’Riordan, CM and others},
  journal={Space Science Reviews},
  volume={220},
  number={5},
  pages={58},
  year={2024},
  publisher={Springer}
}

@article{ellis2010gravitational,
  title={Gravitational lensing: a unique probe of dark matter and dark energy},
  author={Ellis, Richard S},
  journal={Philosophical Transactions of the Royal Society A: Mathematical, Physical and Engineering Sciences},
  volume={368},
  number={1914},
  pages={967--987},
  year={2010},
  publisher={The Royal Society}
}

@article{will2014confrontation,
  title={The confrontation between general relativity and experiment},
  author={Will, Clifford M},
  journal={Living reviews in relativity},
  volume={17},
  number={1},
  pages={1--117},
  year={2014},
  publisher={Springer}
}

@article{sucu2025probing,
  title={Probing Starobinsky-Bel-Robinson gravity: Gravitational lensing, thermodynamics, and orbital dynamics},
  author={Sucu, Erdem and Sakall{\i}, {\.I}zzet},
  journal={Nuclear Physics B},
  volume={1018},
  pages={116982},
  year={2025},
  publisher={Elsevier}
}

@article{einstein1916foundation,
  title={The foundation of the general theory of relativity},
  author={Einstein, Albert and others},
  journal={Annalen Phys},
  volume={49},
  number={7},
  pages={769--822},
  year={1916}
}

@article{wambsganss1998gravitational,
  title={Gravitational lensing in astronomy},
  author={Wambsganss, Joachim},
  journal={Living Reviews in Relativity},
  volume={1},
  number={1},
  pages={12},
  year={1998},
  publisher={Springer}
}

@article{guo2022gravitational,
  title={Gravitational lensing by black holes with multiple photon spheres},
  author={Guo, Guangzhou and Jiang, Xin and Wang, Peng and Wu, Houwen},
  journal={Physical Review D},
  volume={105},
  number={12},
  pages={124064},
  year={2022},
  publisher={APS}
}

@article{fernando2003charged,
  title={Charged black hole solutions in Einstein-Born-Infeld gravity with a cosmological constant},
  author={Fernando, Sharmanthie and Krug, Don},
  journal={General Relativity and Gravitation},
  volume={35},
  number={1},
  pages={129--137},
  year={2003},
  publisher={Springer}
}

@article{crisnejo2018weak,
  title={Weak lensing in a plasma medium and gravitational deflection of massive particles using the Gauss-Bonnet theorem. A unified treatment},
  author={Crisnejo, Gabriel and Gallo, Emanuel},
  journal={Physical Review D},
  volume={97},
  number={12},
  pages={124016},
  year={2018},
  publisher={APS}
}

@article{crisnejo2019higher,
  title={Higher order corrections to deflection angle of massive particles and light rays in plasma media for stationary spacetimes using the Gauss-Bonnet theorem},
  author={Crisnejo, Gabriel and Gallo, Emanuel and Jusufi, Kimet},
  journal={Physical Review D},
  volume={100},
  number={10},
  pages={104045},
  year={2019},
  publisher={APS}
}

@article{lecce2025probing,
  title={Probing axionlike particles with multimessenger observations of neutron star mergers},
  author={Lecce, Francesca and Lella, Alessandro and Lucente, Giuseppe and Vijayan, Vimal and Bauswein, Andreas and Giannotti, Maurizio and Mirizzi, Alessandro},
  journal={Physical Review D},
  volume={112},
  number={2},
  pages={023001},
  year={2025},
  publisher={APS}
}

@article{raffelt1988mixing,
  title={Mixing of the photon with low-mass particles},
  author={Raffelt, Georg and Stodolsky, Leo},
  journal={Physical Review D},
  volume={37},
  number={5},
  pages={1237},
  year={1988},
  publisher={APS}
}

@article{sucu2025scalar,
  title={Scalar-tensor corrections and observational signatures of hairy black holes in horndeski gravity},
  author={Sucu, Erdem and Sakall{\i}, Izzet},
  journal={High Energy Density Physics},
  pages={101220},
  year={2025},
  publisher={Elsevier}
}

@book{perlick2002ray,
  title={Ray Optics, Fermat’s Principle, and Applications to General Relatively},
  author={Perlick, Volker},
  year={2002},
  publisher={Springer}
}

@article{sucu2025quantumRoyal,
  title={Quantum-corrected thermodynamics and plasma lensing of MOG black holes},
  author={Sucu, Erdem and Sakall{\i}, {\.I}zzet},
  journal={Proceedings of the Royal Society A: Mathematical, Physical and Engineering Sciences},
  volume={481},
  number={2320},
  year={2025},
  publisher={The Royal Society}
}

@article{donmez2011development,
  title={On the development of quasi-periodic oscillations in Bondi--Hoyle accretion flows},
  author={D{\"o}nmez, Orhan and Zanotti, Olindo and Rezzolla, Luciano},
  journal={Monthly Notices of the Royal Astronomical Society},
  volume={412},
  number={3},
  pages={1659--1668},
  year={2011},
  publisher={Blackwell Publishing Ltd Oxford, UK}
}

@article{vsramkova2015black,
  title={Black hole spin inferred from 3: 2 epicyclic resonance model of high-frequency quasi-periodic oscillations},
  author={{\v{S}}r{\'a}mkov{\'a}, E and T{\"o}r{\"o}k, G and Kotrlov{\'a}, A and Bakala, P and Abramowicz, MA and Stuchl{\'\i}k, Z and Goluchov{\'a}, K and Klu{\'z}niak, W},
  journal={Astronomy \& Astrophysics},
  volume={578},
  pages={A90},
  year={2015},
  publisher={EDP Sciences}
}

@article{ WOS:001565141800002NPB,
Author = {Sucu, Erdem and Sakalli, Izzet},
Title = {AdS black holes in Einstein-Kalb-Ramond gravity: Quantum corrections,
   phase transitions, and orbital dynamics},
Journal = {NUCLEAR PHYSICS B},
Year = {2025},
Volume = {1018},
Month = {SEP},
DOI = {10.1016/j.nuclphysb.2025.117081},
EarlyAccessDate = {SEP 2025},
Article-Number = {117081},
ISSN = {0550-3213},
EISSN = {1873-1562},
ResearcherID-Numbers = {SAKALLI, İzzet/K-2077-2013
   SUCU, Erdem/JJC-0879-2023},
Unique-ID = {WOS:001565141800002},
}

@article{kruglov2022nonlinearly,
  title={Nonlinearly charged AdS black holes, extended phase space thermodynamics and Joule--Thomson expansion},
  author={Kruglov, SI},
  journal={Annals of Physics},
  volume={441},
  pages={168894},
  year={2022},
  publisher={Elsevier}
}

@article{liang2021joule,
  title={Joule-Thomson expansion of lower-dimensional black holes},
  author={Liang, Jing and Mu, Benrong and Wang, Peng},
  journal={Physical Review D},
  volume={104},
  number={12},
  pages={124003},
  year={2021},
  publisher={APS}
}

@article{akiyama2019first,
  title={First M87 event horizon telescope results. VI. The shadow and mass of the central black hole},
  author={Akiyama, Kazunori and Alberdi, Antxon and Alef, Walter and Asada, Keiichi and Azulay, Rebecca and Baczko, Anne-Kathrin and Ball, David and Balokovi{\'c}, Mislav and Barrett, John and Bintley, Dan and others},
  journal={The Astrophysical Journal Letters},
  volume={875},
  number={1},
  pages={L6},
  year={2019},
  publisher={IOP Publishing}
}

@article{gralla2020measuring,
  title={Measuring the shape of a black hole photon ring},
  author={Gralla, Samuel E},
  journal={Physical Review D},
  volume={102},
  number={4},
  pages={044017},
  year={2020},
  publisher={APS}
}

@article{gralla2019black,
  title={Black hole shadows, photon rings, and lensing rings},
  author={Gralla, Samuel E and Holz, Daniel E and Wald, Robert M},
  journal={Physical Review D},
  volume={100},
  number={2},
  pages={024018},
  year={2019},
  publisher={APS}
}

@article{johnson2020universal,
  title={Universal interferometric signatures of a black hole’s photon ring},
  author={Johnson, Michael D and Lupsasca, Alexandru and Strominger, Andrew and Wong, George N and Hadar, Shahar and Kapec, Daniel and Narayan, Ramesh and Chael, Andrew and Gammie, Charles F and Galison, Peter and others},
  journal={Science advances},
  volume={6},
  number={12},
  pages={eaaz1310},
  year={2020},
  publisher={American Association for the Advancement of Science}
}

@article{johnson2023key,
  title={Key science goals for the next-generation Event Horizon Telescope},
  author={Johnson, Michael D and Akiyama, Kazunori and Blackburn, Lindy and Bouman, Katherine L and Broderick, Avery E and Cardoso, Vitor and Fender, Rob P and Fromm, Christian M and Galison, Peter and G{\'o}mez, Jos{\'e} L and others},
  journal={Galaxies},
  volume={11},
  number={3},
  pages={61},
  year={2023},
  publisher={MDPI}
}

@article{ WOS:001617915000001,
Author = {Sakalli, Izzet and Sucu, Erdem},
Title = {Astrophysical signatures of Konoplya-Zhidenko black holes: Gravitational
   lensing and thermodynamics},
Journal = {INTERNATIONAL JOURNAL OF GEOMETRIC METHODS IN MODERN PHYSICS},
Year = {2025},
Month = {2025 OCT 28},
DOI = {10.1142/S0219887826500404},
EarlyAccessDate = {OCT 2025},
Article-Number = {2650040},
ISSN = {0219-8878},
EISSN = {1793-6977},
ResearcherID-Numbers = {SUCU, Erdem/JJC-0879-2023
   SAKALLI, İzzet/K-2077-2013},
Unique-ID = {WOS:001617915000001},
}

@inproceedings{bambi2018testing,
  title={Testing the Kerr paradigm with the black hole shadow},
  author={Bambi, Cosimo},
  booktitle={The Fourteenth Marcel Grossmann Meeting On Recent Developments in Theoretical and Experimental General Relativity, Astrophysics, and Relativistic Field Theories: Proceedings of the MG14 Meeting on General Relativity, University of Rome “La Sapienza”, Italy, 12--18 July 2015},
  pages={3494--3499},
  year={2018},
  organization={World Scientific}
}

@article{falcke1999viewing,
  title={Viewing the shadow of the black hole at the GalacticCenter},
  author={Falcke, Heino and Melia, Fulvio and Agol, Eric},
  journal={The Astrophysical Journal},
  volume={528},
  number={1},
  pages={L13},
  year={1999},
  publisher={IOP Publishing}
}

@article{bambi2017testing,
  title={Testing the Kerr black hole hypothesis using X-ray reflection spectroscopy},
  author={Bambi, Cosimo and C{\'a}rdenas-Avenda{\~n}o, Alejandro and Dauser, Thomas and Garc{\'\i}a, Javier A and Nampalliwar, Sourabh},
  journal={The Astrophysical Journal},
  volume={842},
  number={2},
  pages={76},
  year={2017},
  publisher={IOP Publishing}
}

@article{sucu2025quantumHassan,
  title={Quantum Corrections in Thermodynamics of Black Holes Modified by Nonlinear Electrodynamics and Their Observational Signatures},
  author={Sucu, Erdem and Sakalli, Izzet and Pourhassan, Behnam},
  journal={International Journal of Geometric Methods in Modern Physics},
  year={2025},
  publisher={World Scientific}
}

@article{davies1989thermodynamic,
  title={Thermodynamic phase transitions of Kerr-Newman black holes in de Sitter space},
  author={Davies, Paul CW},
  journal={Classical and Quantum Gravity},
  volume={6},
  number={12},
  pages={1909},
  year={1989},
  publisher={IOP Publishing}
}

@article{sucu2025quantumOzcan,
  title={Quantum-corrected thermodynamics and plasma lensing in non-minimally coupled symmetric teleparallel black holes},
  author={Sucu, Erdem and Sakall{\i}, Izzet and Sert, {\"O}zcan and Sucu, Yusuf},
  journal={Physics of the Dark Universe},
  pages={102063},
  year={2025},
  publisher={Elsevier}
}

@article{Gibbons:2008rj,
    author = "Gibbons, G. W. and Werner, M. C.",
    title = "{Applications of the Gauss-Bonnet theorem to gravitational lensing}",
    eprint = "0807.0854",
    archivePrefix = "arXiv",
    primaryClass = "gr-qc",
    doi = "10.1088/0264-9381/25/23/235009",
    journal = "Class. Quant. Grav.",
    volume = "25",
    pages = "235009",
    year = "2008"
}

@article{Atamurotov:2021cgh,
    author = {Atamurotov, F. and Jusufi, K. and Jamil, Mv and Abdujabbarov, A. and Azreg-A\"\i{}nou, M.},
    title = "{Axion-plasmon or magnetized plasma effect on an observable shadow and gravitational lensing of a Schwarzschild black hole}",
    eprint = "2109.08150",
    archivePrefix = "arXiv",
    primaryClass = "gr-qc",
    doi = "10.1103/PhysRevD.104.064053",
    journal = "Phys. Rev. D",
    volume = "104",
    number = "6",
    pages = "064053",
    year = "2021"
}

@article{Mendonca:2019eke,
    author = "Mendon\c{c}a, J. T. and Rodrigues, J. D. and Ter\c{c}as, H.",
    title = "{Axion production in unstable magnetized plasmas}",
    eprint = "1901.05910",
    archivePrefix = "arXiv",
    primaryClass = "physics.plasm-ph",
    doi = "10.1103/PhysRevD.101.051701",
    journal = "Phys. Rev. D",
    volume = "101",
    number = "5",
    pages = "051701",
    year = "2020"
}

@article{Wilczek:1987mv,
    author = "Wilczek, F.",
    title = "{Two Applications of Axion Electrodynamics}",
    reportNumber = "NSF-ITP-86-147",
    doi = "10.1103/PhysRevLett.58.1799",
    journal = "Phys. Rev. Lett.",
    volume = "58",
    pages = "1799",
    year = "1987"
}

@book{Synge:1960ueh,
    editor = "Synge, J. L.",
    title = "{Relativity: The General theory}",
    year = "1960"
}

@article{bozza2002gravitational,
  title={Gravitational lensing in the strong field limit},
  author={Bozza, V.},
  journal={Physical Review D},
  volume={66},
  number={10},
  pages={103001},
  year={2002},
  publisher={APS}
}

@article{EventHorizonTelescope:2019dse,
    author = "Akiyama, K. and others",
    collaboration = "Event Horizon Telescope",
    title = "{First M87 Event Horizon Telescope Results. I. The Shadow of the Supermassive Black Hole}",
    eprint = "1906.11238",
    archivePrefix = "arXiv",
    primaryClass = "astro-ph.GA",
    doi = "10.3847/2041-8213/ab0ec7",
    journal = "Astrophys. J. Lett.",
    volume = "875",
    pages = "L1",
    year = "2019"
}

@article{Ayon-Beato:1999qin,
    author = "Ayon-Beato, Eloy and Garcia, Alberto",
    title = "{Nonsingular charged black hole solution for nonlinear source}",
    eprint = "gr-qc/9911084",
    archivePrefix = "arXiv",
    doi = "10.1023/A:1026640911319",
    journal = "Gen. Rel. Grav.",
    volume = "31",
    pages = "629--633",
    year = "1999"
}

@article{Born:1934gh,
    author = "Born, M. and Infeld, L.",
    title = "{Foundations of the new field theory}",
    doi = "10.1098/rspa.1934.0059",
    journal = "Proc. Roy. Soc. Lond. A",
    volume = "144",
    number = "852",
    pages = "425--451",
    year = "1934"
}

@article{Kruglov:2015fbl,
    author = "Kruglov, S. I.",
    title = "{Universe acceleration and nonlinear electrodynamics}",
    eprint = "1601.06309",
    archivePrefix = "arXiv",
    primaryClass = "gr-qc",
    doi = "10.1103/PhysRevD.92.123523",
    journal = "Phys. Rev. D",
    volume = "92",
    number = "12",
    pages = "123523",
    year = "2015"
}

@article{Javed:2025dit,
    author = "Javed, Faisal and Zeeshan Gul, M. and Donmez, Orhan and Naseer, Tayyab and Alshehri, Mansoor H.",
    title = "{Joule-Thomson expansion with Barrow entropy and particle dynamics of charged Rastall-AdS black hole}",
    doi = "10.1016/j.nuclphysb.2025.117001",
    journal = "Nucl. Phys. B",
    volume = "1018",
    pages = "117001",
    year = "2025"
}

@article{Mustafa:2024fau,
    author = {Mustafa, G. and Javed, Faisal and Maurya, S. K. and Alkarni, Shalan and Donmez, Orhan and Cilli, Arzu and G{\"u}dekli, Ertan},
    title = "{Joule-Thomson expansion, motion of particles and QPOs around Bardeen-AdS black hole immersed in a fluid of strings}",
    doi = "10.1016/j.jheap.2024.10.017",
    journal = "JHEAp",
    volume = "44",
    pages = "437--456",
    year = "2024"
}

@article{Donmez:2022dze,
    author = "Donmez, Orhan and Dogan, Fatih and Sahin, Tugba",
    title = "{Study of Asymptotic Velocity in the Bondi{\textendash}Hoyle Accretion Flows in the Domain of Kerr and 4-D Einstein{\textendash}Gauss{\textendash}Bonnet Gravities}",
    eprint = "2205.14382",
    archivePrefix = "arXiv",
    primaryClass = "astro-ph.HE",
    doi = "10.3390/universe8090458",
    journal = "Universe",
    volume = "8",
    number = "9",
    pages = "458",
    year = "2022"
}

@article{hawking1974black,
  title={Black hole explosions?},
  author={Hawking, Stephen W},
  journal={Nature},
  volume={248},
  number={5443},
  pages={30--31},
  year={1974},
  publisher={Nature Publishing Group UK London}
}

@article{sucu2024effect,
  title={The effect of quark--antiquark confinement on the deflection angle by the NED black hole},
  author={Sucu, Erdem and {\"O}vg{\"u}n, Ali},
  journal={Physics of the Dark Universe},
  volume={44},
  pages={101446},
  year={2024},
  publisher={Elsevier}
}

@article{o2024cosmology,
  title={Cosmology of axion dark matter},
  author={O'Hare, Ciaran AJ},
  journal={arXiv preprint arXiv:2403.17697},
  year={2024}
}

@article{sushkov2023quantum,
  title={Quantum science and the search for axion dark matter},
  author={Sushkov, Alexander O},
  journal={PRX Quantum},
  volume={4},
  number={2},
  pages={020101},
  year={2023},
  publisher={APS}
}

@article{bisnovatyi2010gravitational,
  title={Gravitational lensing in a non-uniform plasma},
  author={Bisnovatyi-Kogan, GS and Tsupko, O Yu},
  journal={Monthly Notices of the Royal Astronomical Society},
  volume={404},
  number={4},
  pages={1790--1800},
  year={2010},
  publisher={Blackwell Publishing Ltd Oxford, UK}
}

@article{sucu2025exploring,
  title={Exploring Lorentz-violating effects of Kalb-Ramond field on charged black hole thermodynamics and photon dynamics},
  author={Sucu, Erdem and Sakallı, {\.I}zzet},
  journal={Physical Review D},
  volume={111},
  number={6},
  pages={064049},
  year={2025},
  publisher={APS}
}

@article{sucu2025nonlinear,
  title={Nonlinear electrodynamics effects on the geometry, thermodynamics, and quantum dynamics of (2+ 1)-dimensional black holes},
  author={Sucu, Erdem and Sakall{\i}, Izzet},
  journal={Nuclear Physics B},
  volume={116894},
  year={2025},
  publisher={Elsevier}
}

@article{Donmez:2025piv,
    author = "Donmez, O.",
    title = "{Accretion dynamics and QPO signatures around quantum-corrected black hole: a comparison with Kerr spacetime}",
    doi = "10.1140/epjc/s10052-025-14779-6",
    journal = "Eur. Phys. J. C",
    volume = "85",
    number = "9",
    pages = "1019",
    year = "2025"
}

@article{Donmez:2024gqb,
    author = "Donmez, Orhan",
    title = "{From low- to high-frequency QPOs around the non-rotating hairy Horndeski black hole: Microquasar GRS 1915+105}",
    eprint = "2408.10102",
    archivePrefix = "arXiv",
    primaryClass = "astro-ph.HE",
    doi = "10.1016/j.jheap.2024.11.002",
    journal = "JHEAp",
    volume = "45",
    pages = "1--18",
    year = "2025"
}

@article{Donmez:2024luc,
    author = "Donmez, Orhan and Dogan, Fatih",
    title = "{Estimating the possible QPOs of M87{\ensuremath{*}} from the parameters of a hairy Kerr black hole}",
    eprint = "2407.01478",
    archivePrefix = "arXiv",
    primaryClass = "gr-qc",
    doi = "10.1016/j.dark.2024.101718",
    journal = "Phys. Dark Univ.",
    volume = "46",
    pages = "101718",
    year = "2024"
}

@article{Donmez:2024eff,
    author = "Donmez, Orhan",
    title = "{The comparison of alternative spacetimes using the spherical accretion around the black hole}",
    eprint = "2405.15467",
    archivePrefix = "arXiv",
    primaryClass = "gr-qc",
    doi = "10.1142/S0217732324500767",
    journal = "Mod. Phys. Lett. A",
    volume = "39",
    number = "16",
    pages = "2450076",
    year = "2024"
}

@article{Koyuncu:2014nga,
    author = {Koyuncu, Fahrettin and D{\"o}nmez, Orhan},
    title = "{Numerical simulation of the disk dynamics around the black hole: Bondi Hoyle accretion}",
    doi = "10.1142/S0217732314501156",
    journal = "Mod. Phys. Lett. A",
    volume = "29",
    pages = "1450115",
    year = "2014"
}

@article{Donmez:2010sx,
    author = "Donmez, Orhan",
    title = "{On the development of the Papaloizou{\textendash}Pringle instability of the black hole{\textendash}torus systems and quasi-periodic oscillations}",
    eprint = "1304.0584",
    archivePrefix = "arXiv",
    primaryClass = "astro-ph.HE",
    doi = "10.1093/mnras/stt2255",
    journal = "Mon. Not. Roy. Astron. Soc.",
    volume = "438",
    number = "1",
    pages = "846--858",
    year = "2014" 
}

@article{Mustafa:2025mkc,
    author = "Mustafa, G. and Ghosh, Sushant G. and Donmez, Orhan and Maurya, S. K. and Orzuev, Shakhzod and Atamurotov, Farruh",
    title = "{Testing quantum-corrected black holes with QPOs observations: a study of particle dynamics and accretion~flow}",
    eprint = "2506.16405",
    archivePrefix = "arXiv",
    primaryClass = "gr-qc",
    doi = "10.1088/1475-7516/2025/10/068",
    journal = "JCAP",
    volume = "10",
    pages = "068",
    year = "2025"
}

@article{Mustafa:2025gjv,
    author = "Mustafa, G. and Javed, Faisal and Maurya, S. K. and Ditta, A. and Donmez, Orhan and Naseer, Tayyab and Bouzenada, Abdelmalek and Atamurotov, Farruh",
    title = "{Magnetized particle motion and accretion process with shock cone morphology around a decoupled hairy black holes}",
    eprint = "2511.17137",
    archivePrefix = "arXiv",
    primaryClass = "astro-ph.HE",
    month = "11",
    year = "2025"
}

@article{Zhang:2022epl,
    author = "Zhang, Yuexin and M{\'e}ndez, Mariano and Garc{\'\i}a, Federico and Karpouzas, Konstantinos and Zhang, Liang and Liu, Honghui and Belloni, Tomaso M. and Altamirano, Diego",
    title = "{The evolution of the high-frequency variability in the black hole candidate GRS~1915+105 as seen by RXTE}",
    eprint = "2204.05273",
    archivePrefix = "arXiv",
    primaryClass = "astro-ph.HE",
    doi = "10.1093/mnras/stac1050",
    journal = "Mon. Not. Roy. Astron. Soc.",
    volume = "514",
    number = "2",
    pages = "2891--2901",
    year = "2022"
}

@article{Belloni:2019sot,
    author = "Belloni, Tomaso M. and Bhattacharya, Dipankar and Caccese, Pietro and Bhalerao, Varun and Vadawale, Santosh and Yadav, J. S.",
    title = "{A variable-frequency HFQPO in GRS 1915+105 as observed with AstroSat}",
    eprint = "1908.00437",
    archivePrefix = "arXiv",
    primaryClass = "astro-ph.HE",
    doi = "10.1093/mnras/stz2143",
    journal = "Mon. Not. Roy. Astron. Soc.",
    volume = "489",
    number = "1",
    pages = "1037--1043",
    year = "2019"
}

@article{Stuchlik:2016brp,
    author = "Stuchl{\'\i}k, Zden{\v{e}}k and Kolo{\v{s}}, Martin",
    title = "{Controversy of the GRO J1655-40 black hole mass and spin estimates and its possible solutions}",
    eprint = "1608.01659",
    archivePrefix = "arXiv",
    primaryClass = "astro-ph.HE",
    doi = "10.3847/0004-637X/825/1/13",
    journal = "Astrophys. J.",
    volume = "825",
    pages = "13",
    year = "2016"
}

@article{Chakrabarti:2005bd,
    author = "Chakrabarti, Sandip K. and Nandi, A. and Debnath, D. and Sarkar, R. and Datta, B. G.",
    title = "{Propagating oscillatory shock model for QPOs in GRO J1655-40 during the March 2005 outburst}",
    eprint = "astro-ph/0508024",
    archivePrefix = "arXiv",
    journal = "Indian J. Phys. B",
    volume = "78",
    pages = "1--5",
    year = "2005"
}

@article{Wang:2008xda,
    author = "Wang, Ding-Xiong and Gan, Zhao-Ming and Huang, Chang-Yin and Li, Yang",
    title = "{Association of the 3:2 HFQPO Pairs with the Broad Fe K Line in XTE J1550-564 and GRO J1655-40}",
    eprint = "0809.2695",
    archivePrefix = "arXiv",
    primaryClass = "astro-ph",
    doi = "10.1111/j.1365-2966.2008.13965.x",
    journal = "Mon. Not. Roy. Astron. Soc.",
    volume = "391",
    pages = "1332",
    year = "2008"
}
\bibliographystyle{apsrev}
\end{document}